\documentclass [11pt]{article}
\pdfoutput=1
\usepackage{amsfonts}
\usepackage{graphicx}
\usepackage{amsmath,amssymb, dsfont}
\usepackage{latexsym}
\usepackage{mathrsfs}
\usepackage{bbold}
\usepackage{color}
\usepackage{slashed}
\usepackage{cancel}
\usepackage{soul} %
\usepackage{setspace} %
\usepackage{comment}
\usepackage{booktabs}
\usepackage{multirow}
\usepackage{cite}
\usepackage{tabularx}
\usepackage[small]{caption}
\usepackage{subcaption}
\usepackage{tikz}
\usetikzlibrary{decorations.pathreplacing}
\usepackage{scalerel}

\makeatletter
\def\smalloverbrace#1{%
  \@ifnextchar^{\tikz@@overbrace{#1}}{\tikz@@overbrace{#1}^{}}}
\def\tikz@@overbrace#1^#2{%
  \tikz[baseline=(a.base)] {\node[inner sep=0] (a) {\(#1\)};
  \draw[line cap=round,decorate,decoration={brace,amplitude=5pt}]
    (a.north west) -- node[pos=.5,above,inner sep=7pt] {\(\scriptstyle #2\)} (a.north east);}}
\makeatother

\definecolor{red}{rgb}{1,0,0}

\makeatletter
\def\section{\@startsection {section}{1}{\z@}{-3.5ex plus -1ex minus
 -.2ex}{2.3ex plus .2ex}{\large\bf}}
\def\subsection{\@startsection{subsection}{2}{\z@}{-3.25ex plus -1ex
minus -.2ex}{1.5ex plus .2ex}{\normalsize\bf}}
\makeatother
\makeatletter

\@addtoreset{equation}{section}

\makeatother

\newcommand{\bea}{\begin{equation} \begin{aligned}} \newcommand{\eea}{\end{aligned} \end{equation}}
\def\be{\begin{equation}} \def\ee{\end{equation}} 
\def\nn{\nonumber}
\def\beaa{\begin{eqnarray}} \def\eeaa{\end{eqnarray}}
\def\Dslash{\raisebox{1pt}{$\slash$} \hspace{-8pt} D}

\usepackage[colorlinks,linkcolor=black,citecolor=blue,urlcolor=blue,linktocpage,pagebackref]{hyperref}

\allowdisplaybreaks

\begin{document}

\thispagestyle{empty}

\begin{center}

	\vspace*{-.6cm}

	\begin{center}

		\vspace*{1.1cm}

		{\centering \Large\textbf{Cancellation of IR Divergences in 3d Abelian Gauge Theories}}

	\end{center}

	\vspace{0.8cm}
	{\bf Giovanni Galati and Marco Serone}

	\vspace{1.cm}

	{\em SISSA, Via Bonomea 265, I-34136 Trieste, Italy}

	\vspace{.3cm}

	{\em INFN, Sezione di Trieste, Via Valerio 2, I-34127 Trieste, Italy}

	\vspace{.3cm}

\end{center}

\vspace{1cm}

\centerline{\bf Abstract}
\vspace{2 mm}
\begin{quote}

Three dimensional abelian gauge theories classically in a Coulomb phase 
are affected by IR divergences even when the matter fields are all massive.
Using generalizations of Ward-Takahashi identities, we show that correlation functions of gauge-invariant 
operators are IR finite to all orders in perturbation theory. 
Gauge invariance is sufficient but not necessary for IR finiteness.  In particular we show that specific gauge-variant correlators, 
including the two-point function of matter fields, are also IR finite to all orders in perturbation theory. 
Possible applications of these results are briefly discussed.
\end{quote}

\newpage

\tableofcontents

\section{Introduction}

Three dimensional (3d) abelian gauge theories are interesting for a variety of reasons. Among several other applications,
 the bosonic theory describes the effective physics of ordinary superconductors near the phase transition \cite{Ginzburg:1950sr}, 
both the fermionic (QED$_{3}$) and the bosonic (sQED$_{3}$) theories describe quantum phase transitions in certain anti-ferromagnetic 
spin lattice systems \cite{doi:10.1126/science.1091806,PhysRevB.70.144407}, and QED can play a role in the physics of high-$T_c$ cuprate superconductors \cite{Rantner:2000wer}. 
From a theoretical point of view abelian 3d gauge theories have also been shown to enjoy interesting duality relations \cite{Karch:2016sxi,Seiberg:2016gmd}.

 Being strongly coupled in the IR, a basic fundamental issue is to understand whether the vacuum is gapped or gapless, and the nature of the low-energy
 fluctuations. Three dimensional theories, however, suffer from severe IR divergences that hinder a direct quantitative investigation in three space-time dimensions.
 A reliable known way to tame IR divergences is to take a large $N$ limit, where $N$ is the number of matter fields.\footnote{A precursor of this observation dates back to \cite{Jackiw:1980kv}. There it was argued that IR divergences in the massless case for any $N$ can be cured by non analytic terms in the coupling constant which arise when an infinite class of diagrams is resummed using certain gap equations. Evidence for the correcteness of this proposal has been recently provided in a certain $d=3$ scalar theory on a  lattice \cite{Cossu:2020yeg}.} In this limit it has been shown that both QED$_{3}$ \cite{Appelquist:1986qw,Appelquist:1988sr,Nash:1989xx}  and sQED$_3$ \cite{Appelquist:1981sf,Appelquist:1981vg} flow in the IR to a non-trivial CFT. The fixed point is expected to persist at finite $N$ up to some (yet to be determined) critical value $N_c$. In addition to $N$, in sQED$_3$ the nature of the phase transition could depend on the so called Ginzburg parameter $k = e^2/\lambda$, where $e^2$ is the gauge coupling and $\lambda$ is the scalar quartic coupling. More precisely, there could be separatrix lines delimiting different RG flows. Depending on the initial UV values of $k$, the theory could then either flow to a fixed point, or to another one, or end up in a first-order phase transition. At large $N$ no separatrix lines appear, but they could be present at finite $N$.\label{footnoteK}
  
Another way to get rid of IR divergences and directly access the critical theory is obtained using the $\epsilon$-expansion starting from $d=4-\epsilon$ dimensions. 
A perturbative analysis predicts for sQED$_{3}$
a fixed point for $N> N_\epsilon$ \cite{Halperin:1973jh}, where the value of $N_\epsilon$ sensitively depends on the loop order \cite{Ihrig:2019kfv}. 
A similar conclusion has been reached more recently for QED$_3$ in \cite{DiPietro:2015taa} and, using also the $F$-theorem, in \cite{Giombi:2015haa}.

Large $N$ and $\epsilon$-expansion techniques are not sufficient to study the theory at small $N$ and at $d=3$ which are in fact the cases of more physical interest (for instance the superconductor physics corresponds to $N = 1$). This regime is however accessible by numerical lattice simulations. They predict for $N=1$ sQED$_3$ a first-order phase transition when the Ginzburg parameter $k$ is sufficiently small and a second order one for large $k$ 
(see e.g. \cite{Dasgupta:1981zz,Bartholomew:1983zz,Kajantie:1997vc}). The presence of a second-order phase transition for sufficiently large $k$ is in fact guaranteed by
particle-vortex duality \cite{Peskin:1977kp,Dasgupta:1981zz}.
For QED$_3$ the estimates of $N_c$ under which chiral symmetry breaking occurs was predicted to be $N_c=1$ \cite{Hands:2002dv,Hands:2004bh} while recent lattice simulations claim that $N_c=0$ \cite{Karthik:2016ppr}. 
Another first principle approach is the conformal bootstrap, which allows to put general bounds on the properties of the critical theory and can rigorously rule out 
disallowed scenarios. See section V.E of \cite{Poland:2018epd} for an overview of the results obtained in this way on $3d$ abelian gauge theories.
Functional renormalization group methods can also be used in order to understand the superconducting phase diagram, where IR divergences are regularized by the introduction of an infrared cut-off. The qualitative picture found is in line with the one found by lattice simulations and described above \cite{Bergerhoff:1995zq}.

As we mentioned, IR divergences do not allow to study the theory perturbatively when matter and gauge fields are massless in the UV. 
On the other hand,  we could consider the theory off-criticality by giving mass to matter fields. The situation is similar to that in quartic scalar models, which can be studied at criticality using large $N$ or $\epsilon$-expansion, or off-criticality at $d=3$ by Borel resumming the perturbative series \cite{Parisi:1980gya}.\footnote{Borel resummation is also needed in the $\epsilon$-expansion if one wants to reliably reach $d=3$.}
The presence of a massless photon in abelian gauge theories, however, does not guarantee that IR divergences are all gone. 
As well-known, in $4d$ Lorentzian abelian gauge theories IR divergences cancel in cross-sections where a sum over amplitudes with external 
soft photons is included \cite{Bloch:1937pw}. As mentioned, $3d$ abelian gauge theories are strongly coupled in the IR and before worrying about
how IR divergences possibly cancel in a putative gapless phase, we should understand which are the degrees of freedom and how they interact.
In Euclidean space, however, the observables are not cross-sections, but (among others) correlation functions of local operators, and the question of IR finiteness can be posed.
We simply have to require that such correlation functions should be IR finite.

\begin{figure}[t!]
 \centering
   \raisebox{-0 em}{\includegraphics[width=10cm]{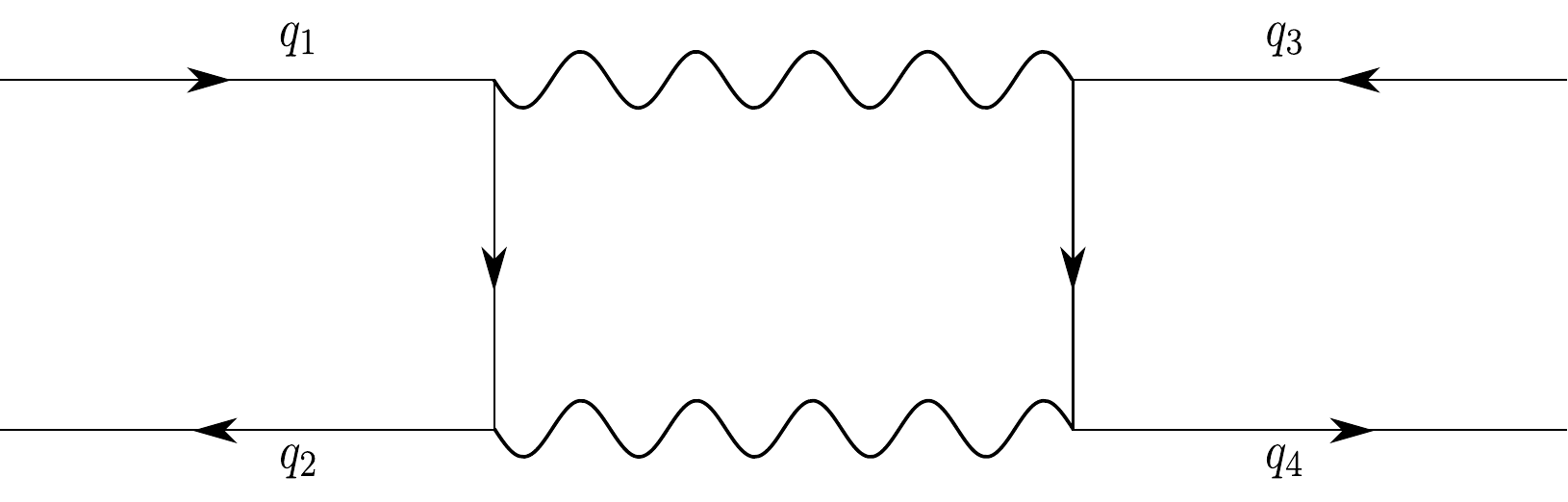}}
    \caption{Example of a one-loop IR divergent diagram in a four-fermion correlator in QED$_3$. The momenta are all incoming with $q_1+q_2=0= - (q_3+q_4)$.}
  \label{fig:4fermion}
\end{figure}

While typically correlation functions are IR regulated by the finite momenta of the external operators, IR divergences can appear 
order by order in perturbation theory when a partial sum of the external momenta sum up to zero. Such configurations are called exceptional.
A simple example is provided by the one-loop correction to the four-point matter correlator in both QED$_3$ and sQED$_3$, when the external momenta have exceptional momenta.
See fig.\ref{fig:4fermion} for an illustration in QED$_3$.
Such IR divergences can hinder a perturbative study of these theories.\footnote{In sQED$_3$, for example, IR divergences did not allow in \cite{Herbut:1996ut} to fix the quartic scalar coupling without introducing new parameters which cannot be fixed from first principles.}

The aim of this paper is to prove that to all orders in perturbation theory correlation functions of gauge-invariant operators in Euclidean abelian gauge theories with massive matter are IR finite. This result applies independently of whether the abelian gauge theory is an effective description or a UV-complete fundamental theory.

We start in section \ref{sec:IRFG} by showing our proof, based on simple generalizations of Ward-Takahashi identities and the properties of certain effective vertices obtained when matter is integrated out \cite{Coleman:1985zi}.\footnote{We are surprised that, to our knowledge,  a proof like ours did not appear before in the literature, given its simplicity. 
On the other hand,  the study of the IR properties of correlators in $d<4$ field theories at finite $N$ and fixed dimension does not seem to have received 
much attention in the literature. One exception is \cite{David:1980rr} where, following observations in \cite{Jevicki:1977zn,Elitzur:1978ww}, it was shown
that correlation functions of $O(N)$-invariant observables in $2d$ non-linear $O(N)$ sigma models are IR finite in the naive vacuum  
where $O(N)$ is spontaneously broken and Goldstone bosons appear (as well-known, this is not the actual quantum vacuum of the theory, 
given that the breaking of continuous global symmetries is forbidden in $d=2$ \cite{Mermin:1966fe,Coleman:1973ci}).}
In section \ref{sec:Example} we give two examples of 2-point functions of gauge-invariant operators up to some loop order and illustrate how the 
proof in section \ref{sec:IRFG} works in explicit cases.  Gauge-invariance is actually not necessary to get IR finite results. We focus in section \ref{sec:gauge_variant} on a notable example of this sort, the two-point function of elementary matter fields, and show that this non-gauge invariant correlator is IR finite to all orders in perturbation theory.  Our perturbative results are insensitive to the global structure of the gauge group, while non-perturbative effects, related to monopole operators, depend on that.
In order to clarify in which regimes euclidean perturbative correlators are expected to be not affected by monopole effects, we briefly review in section \ref{sec:moop} the role of monopole operators. We give an outlook of possible applications of our results in section \ref{sec:outlook}.

\section{IR finiteness of gauge invariant correlation functions}
\label{sec:IRFG} 

Consider a 3d Euclidean abelian gauge theory coupled to matter. The latter can be made of scalars or fermions, or both.
Chern-Simons terms play an important role in 3d gauge theories, but
from our point of view they are ``trivial" since they provide a mass to the photon
and automatic IR finiteness. We then assume a parity symmetry and an even number of fermions with parity invariant mass terms in the theory, so that no Chern-Simons terms are allowed. We assume that all matter fields are massive and that the theory is classically in a Coulomb phase with a massless photon.
The gauge theory can be an effective field theory description of some microscopic theory, such as a spin system on a lattice where the gauge field is possibly emergent, or it can be a fundamental UV-complete theory. Let us denote by $S$ the total action, sum of the gauge, matter, and gauge fixing terms:
\be
S =S_\gamma+S_M+ S_{g.f.} \,,
\label{eq:Stot}
\ee
where
\beaa
S_\gamma(A)  & = &   -\frac{1}{4}  \int \! d^d x \, F_{\mu\nu}^2+\ldots  \,, \quad \quad  S_{g.f.}(A)  =  \frac{1}{2\xi}  \int \! d^d x \,(\partial_\mu A^\mu)^2\,,
\label{eq:SgammaGF}  \\
 S_M(A_\mu, \psi, \phi) & =& \sum_{i=1}^{2N_f} \bar \psi_i (i\Dslash - m_i) \psi_i + \sum_{j=1}^{N_s} \Big( |D \phi_j|^2 + m_j^2 |\phi_j|^2\Big)  +\ldots \,.
\label{eq:SM}
\eeaa
In \eqref{eq:SM}, the $\psi_i$ are two-component fermions and the fermion masses $m_i$ are such that parity is preserved. The $\ldots$ in \eqref{eq:SgammaGF} and \eqref{eq:SM} denote possible higher dimensional operators in the effective field theory description. In \eqref{eq:SM} they include 
possible self-interactions among the scalar fields $\phi_j$, among scalars and fermions, Lagrange multipliers enforcing constraints among the scalars, like in $\mathbb{CP}^{N}$ models, etc. In fact, the matter action is quite arbitrary, as long as matter fields are massive and $U(1)$ is linearly realized. 
QED$_3$ and sQED$_3$ correspond of course to  $N_s=0$ and $N_f=0$ in \eqref{eq:SM}, respectively. 

We would like to show that arbitrary correlation functions of gauge-invariant operators based on the action \eqref{eq:Stot} are IR finite.
Our main argument is based on manipulations very similar to those used in \cite{Coleman:1985zi} to prove that the Chern-Simons level in an abelian gauge theory coupled to massive matter, beyond one-loop level, does not receive further corrections to all orders in perturbation theory.
We first define an effective action $S_{eff}(A)$ for the photon field obtained by integrating out the massive matter degrees of freedom:
\be
e^{-S_{eff}(A)} = e^{-S_\gamma(A)} \int\!{\cal D}\Phi \, e^{-S_M(\Phi,A)}\,.
\label{eq:Seff2}
\ee
In \eqref{eq:Seff2} we collectively denote by $\Phi$ any (bosonic and fermionic)  matter field in the theory.
Since $S_{eff} $ is gauge-invariant, under an infinitesimal $U(1)$ transformation we get
\be
\partial_\mu \frac{\delta S_{eff}}{\delta A_\mu} = 0\,.
\label{eq:SeffGI}
\ee
Taking functional derivatives with respect to $A_\mu(x_i)$ $n-1$ times give us in momentum space the relations
\be
p^{\mu_i}_i \gamma^{(n)}_{\mu_1\ldots \mu_n}(p_1,\ldots, p_n) = 0\,, \quad \forall i=1,\ldots, n\,,
\label{eq:WIforgamma}
\ee
where $\gamma^{(n)}$ are the Fourier transforms\footnote{Here and in what follows, with an abuse of language we will denote a function and its Fourier transform with the same symbol.} of the 1PI $n$-point functions for {\it non-dynamical} photons:
\be
\gamma_{\mu_1 \ldots \mu_n}^{(n)}(y_j) \equiv \bigg(\prod_{j=1}^n \frac{\delta}{\delta A_{\mu_j}(y_j)} \bigg) S_{eff}(A)  \Big|_{A=0}\,.
\label{eq:gammaDef}
\ee
While $\gamma^{(2)}$ represents the tree-level photon propagator including matter corrections,  
$\gamma^{(n)}$ with $n>2$ represents effective vertices in the low energy photon effective field theory. To all orders in perturbation theory, the 1PI $m$-photon 
amplitudes $G_\gamma^{(m)}$ are obtained by gluing in all possible ways all the vertices $\gamma^{(n)}$, with $n=3,4,\ldots$ through effective photon lines, constructing in this way all possible Feynman diagrams. Crucially, when matter is massive, the functions $\gamma^{(n)}$ are {\it analytic} at the origin in momentum space individually for each $p_i$.
In this case it is simple to show that the $\gamma^{(n)}$'s have to vanish whenever any momentum $p_i=0$.
Let us consider $i=1$ in \eqref{eq:WIforgamma} and take a derivative with respect to $p_1^\nu$:
\be
\gamma_{\nu \mu_2\ldots \mu_n}^{(n)}(p_1,\ldots,p_n) + p_1^\mu  \frac{\partial \gamma_{\mu_1 \ldots \mu_n}^{(n)}(p_1,\ldots,p_n)}{\partial p_1^\nu} = 0\,.
\label{eq:CH1}
\ee
Since $\gamma^{(n)}$ are analytic functions of the momenta, the derivative $\partial_\nu \gamma^{(n)}$ appearing in the second term of \eqref{eq:CH1} is finite.
Hence, when $p_1^\mu\rightarrow 0$, \eqref{eq:CH1} implies that
\be
\gamma_{\mu_1 \mu_2\ldots \mu_n}^{(n)}(0,p_2, \ldots,p_n) =0\,.
\label{eq:CH2}
\ee 
Analyticity implies also that for small $p_1$, $\gamma_{\mu_1 \mu_2\ldots \mu_n}^{(n)} = O(p_1)$. The argument can be repeated for the other $p_k$'s. 
Since we have $n-1$ independent momenta we get $\gamma_{\mu_1 \mu_2\ldots \mu_n}^{(n)} = O(p_1 \ldots p_{n-1})$.
Using Bose symmetry and Lorentz invariance the argument can be improved to include $p_n$ \cite{Coleman:1985zi}. 
In this way we finally get that for small $p_i$'s 
\be
\gamma^{(n)}_{\mu_1\ldots \mu_n}(p_1,\ldots, p_n) = O(p_1 \ldots p_n)\,.
\label{eq:gammapzero}
\ee
To all orders in perturbation theory, for small momentum $p$ the effective photon propagator goes like the tree-level one, $\propto 1/p^2$.\footnote{Beyond perturbation theory this is no longer true. For instance, at large $N$ the photon propagator goes like $1/p$ for small momenta.} Any internal photon line has to attach to a pair of $\gamma^{(n)}$'s or to the same $\gamma^{(n)}$. In both cases the vertices bring two powers of $p$, precisely canceling the $1/p^2$ factor for each photon line. The IR finiteness of $G_\gamma^{(m)}$ is then proved.

\begin{figure}[t!]
 \centering
   \raisebox{-0 em}{\includegraphics[width=10cm]{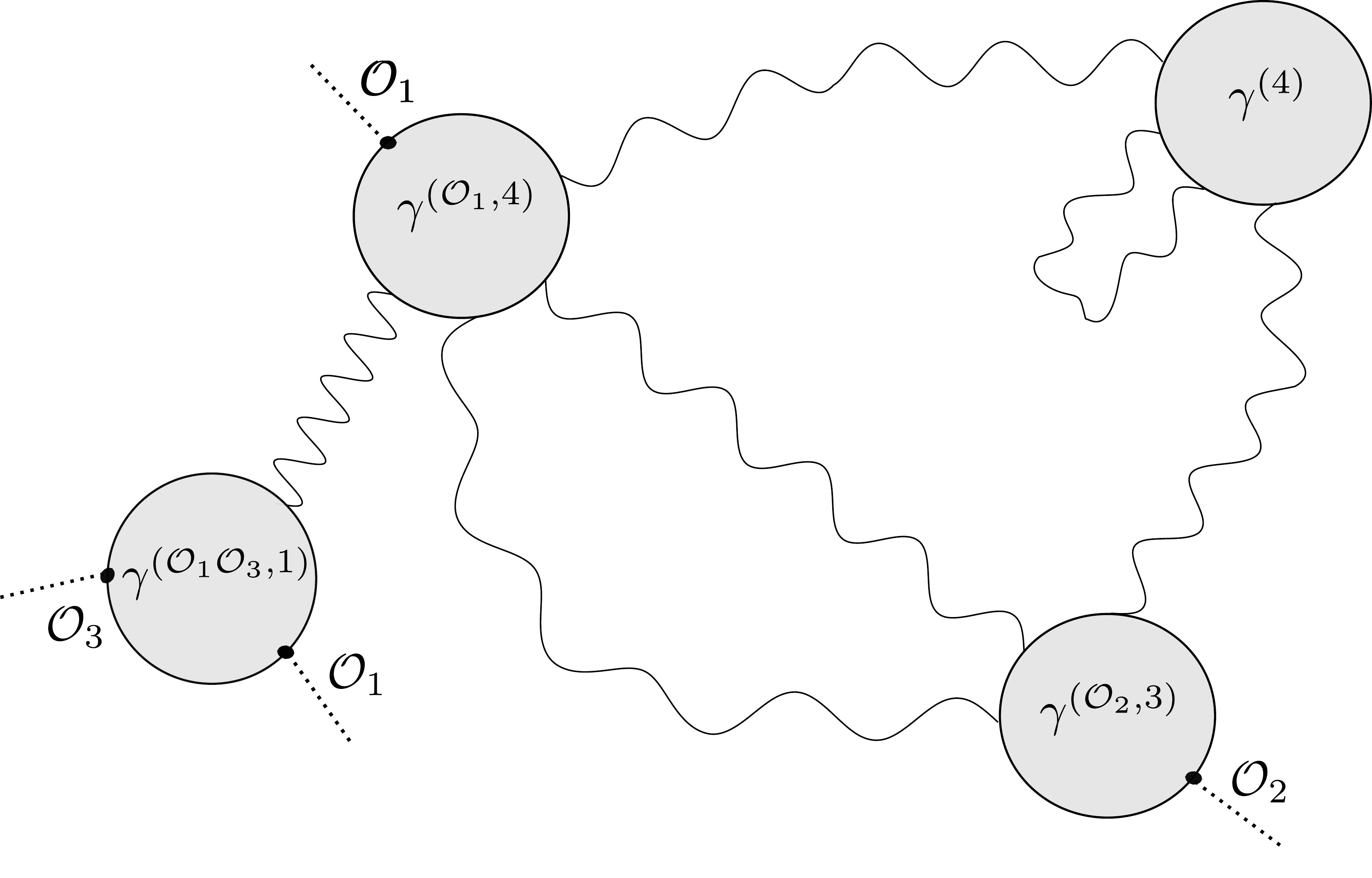}}
    \caption{Example of diagram with four effective vertices contributing to the connected four-point function $\langle \mathcal{O}_1 \mathcal{O}_1 \mathcal{O}_2 \mathcal{O}_3\rangle$. When $\mathcal{O}_i$ are gauge-invariant operators, the vertices provide powers of momenta which cancel the would-be IR singularities coming from the photon propagators.}
  \label{fig:attaching}
\end{figure}

Building on the above argument, we can prove the IR finiteness of arbitrary  correlation functions of gauge-invariant operators $\mathcal{O}_i$ made of matter and/or photon elementary constituents.
Let $J_i$ be sources coupled to the operators $\mathcal{O}_i$. We have for the connected correlator\footnote{In writing \eqref{eq:GIcorrelation} we are assuming that all operators $\mathcal{O}_i$ are distinct.
If not, we obviously have less sources and repeated functional derivatives in \eqref{eq:GIcorrelation}.}
\be
\langle \mathcal{O}_1(x_1) \ldots \mathcal{O}_k (x_k) \rangle = \prod_{i=1}^k \frac{\delta}{\delta J_i(x_i)} W[J_i,J_\mu]\Big|_{J_\mu = J_i=0}\,,
\label{eq:GIcorrelation}
\ee
where
\be
e^{-W[J_i,J_\mu]} = \int\! {\cal D}\Phi{\cal D}A\, e^{-S(\Phi,A) +\int\! d^d x \, ( \sum_{i=1}^k J_i \mathcal{O}_i + J_\mu A_\mu)}\,.
\label{eq:WJiDef}
\ee
As before, we can first define an effective action $S_{eff}$ for the photon field by integrating out the massive matter degrees of freedom:
\be
e^{-S_{eff}(A,J_i)} = e^{-S_\gamma(A)}\int\!{\cal D}\Phi \, e^{-S_M(\Phi,A) +\int\! d^d x \,  \sum_{i=1}^k J_i \mathcal{O}_i}\,.
\label{eq:Seff}
\ee
Crucially, the effective action is now a non-trivial complicated functional of the external currents $J_i$, since $\mathcal{O}_i$ can be made of matter fields. Yet, gauge invariance
guarantees that
\be
\partial_\mu \frac{\delta S_{eff}(A,J_i) }{\delta A_\mu} = 0\,,
\label{eq:SeffGIJi}
\ee
generalization of  \eqref{eq:SeffGI} in presence of the external sources $J_i$. 
We can now define
\be
\gamma_{\mu_1 \ldots \mu_n}^{(\mathcal{O}_1 \ldots \mathcal{O}_k,n)}(x_i,y_j) \equiv \bigg(\prod_{i=1}^k \frac{\delta}{\delta J_i(x_i)} \bigg)\bigg(\prod_{j=1}^n \frac{\delta}{\delta A_{\mu_j}(y_j)} \bigg) S_{eff}(A,J_i)  \Big|_{A=J_i=0}\,.
\label{eq:gammaJiDef}
\ee
In momentum space, \eqref{eq:SeffGIJi} implies that
\be
p_j^{\mu_j} \gamma_{\mu_1 \ldots \mu_n}^{(\mathcal{O}_1 \ldots \mathcal{O}_k,n)}(q_i ,p_j) = 0 \,, \quad \forall j=1,\ldots, n \,,
\label{eq:gammaJiWI}
\ee
where $q_i$ and $p_j$  are the momenta of the composite operators $\mathcal{O}_i$ and of the non-dynamical photons, respectively.
Since the matter is massive, the functions $\gamma^{(\mathcal{O}_1 \ldots \mathcal{O}_k,n)}$ are analytic for $p_j\rightarrow 0$ for {\it arbitrary} values of $q_i$. In particular, 
we can repeat the considerations made below \eqref{eq:WIforgamma} to get for small $p_j$'s\footnote{For $n>1$ all momenta $p_j$ are independent and
\eqref{eq:gammaJi0} follows straightforwardly from \eqref{eq:SeffGI}. This in contrast to \eqref{eq:gammapzero} where Bose symmetry and Lorentz invariance are required to extend $O(p_1 \ldots p_{n-1})$ to $O(p_1\ldots p_n)$ \cite{Coleman:1985zi}.}
\be
\gamma_{\mu_1 \ldots \mu_n}^{(\mathcal{O}_1 \ldots \mathcal{O}_k,n)}(q_i ,p_j) = O(p_{\mu_1} \ldots p_{\mu_n} ) \,.
\label{eq:gammaJi0}
\ee
The full conncected correlator \eqref{eq:GIcorrelation} is obtained by gluing in all possible ways effective vertices $\gamma^{(\mathcal{O}_1 \ldots \mathcal{O}_k,n)}$ through photon lines, see fig. \ref{fig:attaching} for an illustration. If the composite operator carries non-vanishing momentum $q$, this will be carried by some photon leg in the effective vertex. That photon leg would then be $O(q)$ as the virtual momentum goes to zero,
but obviously the photon propagator attached would also be IR regulated by the same $q$.
No matter where photon lines are attached, the potentially dangerous $1/p^2$ factors coming from propagators will either be compensated by
similar factors coming from the effective vertices or IR - regulated by external momenta. The IR finiteness is guaranteed for any value of external momenta,
in particular for any choice of exceptional configuration, including the most IR dangerous configuration obtained when all external fields have vanishing momentum.
It should be emphasized that individual Feynman diagrams can be IR divergent and it is only when summed together that such IR divergences are guaranteed to cancel.

\subsection{Renormalization}

We have so far neglected the effect of UV renormalization, but we now show that no further IR divergences are induced by the renormalization process. This amounts to show that
to each order in perturbation theory diagrams with counterterm insertions are also IR finite. 
In addition to the counterterms required to renormalize the action, we also get the counterterms associated to the composite operators.
Due to operator mixing, these are in general matrix valued. In abelian gauge theories gauge invariant operators can only mix among themselves. This is shown (see e.g. chap. 18 of \cite{zinn2002quantum}) by noting that the solution of the generating functional equation of Ward-Takahashi identities  for the 1PI effective action $\Gamma$ is  
\be
\Gamma = \Gamma_{{\rm GI}} + \frac{1}{2\xi} \int \!\! d^dx (\partial A)^2\,,
\ee
where $\Gamma_{{\rm GI}}$ is a gauge invariant functional made of $A_\mu$, the matter fields, and the external sources $J_i$ associated to the composite operators $\mathcal{O}_i$.
In general the mixing will also involve gauge-invariant redundant operators. As far as IR divergences are concerned, however, the latter do not introduce any complication and can be considered  together with the non-redundant gauge-invariant operators.\footnote{Note that gauge-invariant operators can instead mix with gauge-variant ones in non-abelian gauge theories. Using BRST symmetry, it has been proven \cite{Joglekar:1975nu} (see \cite{Barnich:2000zw} for a more modern perspective in terms of cohomology in a wider context), that gauge-variant operators are always BRST exact and there exists a basis where they decouple.}
As a consequence of this discussion the counter-term action $S_{c.t.}$ will not spoil the gauge invariance of the original action and so the Ward identities (\ref{eq:gammaJi0}) are still valid even for renormalized correlators. Diagrammatically the functions $\gamma^{(\mathcal{O}_1 \ldots \mathcal{O}_k,n)}$, defined considering also $S_{c.t.}$ in (\ref{eq:Seff}), are now the ones in which also counter-term insertions are considered and they can be used as building blocks in order to construct connected {\it renormalized} correlators of gauge-invariant operators. Then, the same arguments used for the bare correlators guarantee the IR finiteness of the renormalized ones.  

We have so far tacitly assumed that the composite operators were Lorentz scalars, but all our considerations are valid for
arbitrary gauge-invariant tensor operators.

For further clarity and illustration, in the next section we will consider two examples of correlators of gauge-invariant operators,
show how the decomposition in terms of the effective vertices \eqref{eq:gammaDef}, \eqref{eq:gammaJiDef} work, and verify the validity of \eqref{eq:gammaJi0} in special cases.

\section{Examples}
\label{sec:Example}

We explicitly verify in this section some of the considerations made before in two specific gauge invariant correlators.
For concreteness we consider the UV complete sQED$_3$ with $N_s=1$ and quartic scalar interactions. 
The first example is the two-point function of the lowest dimensional scalar gauge invariant operator  $\phi^\dagger \phi$, 
where we will show how the decomposition in the building blocks $\gamma$ and the cancellation of IR divergences take place at the first orders in perturbation theory. 
The second example is the two-point function of the tensor operator $F_{\mu \nu} \phi^\dagger \phi$. It enjoys two additional properties with respect to the previous operator: it is a composite operator made of both photon and matter fields and it carries a non trivial Lorentz structure. In this case we want to show that these two additional properties do not spoil the arguments just explained. For simplicity of writing, in both these examples, we will focus only on the 1PI diagrams since connected but non 1PI ones do not bring any further complication regarding IR divergences.
In dealing with one composite operator $\mathcal{O}$ only, it is convenient to introduce a light notation for the building blocks $\gamma^{(\mathcal{O}_1\ldots \mathcal{O}_k,n)}$ introduced in \eqref{eq:gammaJiDef}, and define
\vspace{-0.4cm}
\be
\gamma^{(k,n)} \equiv \gamma^{(\smalloverbrace{\scaleto{\mathcal{O} \ldots \mathcal{O}\mathstrut}{9pt}}^k,n)}\,,
\ee
where $\mathcal{O}=\phi^\dagger \phi$ in subsection \ref{subsec:phiphi} and $\mathcal{O}=F_{\mu \nu}\phi^\dagger \phi$ in subsection \ref{subsec:phiphiF}.

\subsection{$\langle \phi^{\dagger}\phi(q)\phi^{\dagger}\phi(-q)\rangle $} 
\label{subsec:phiphi}

The leading free theory contribution arises at one-loop level and corresponds to $\gamma^{(2,0)}\big\rvert_{\text{1L}}$:
\begin{equation}
  \langle \phi^{\dagger}\phi(q)\phi^{\dagger}\phi(-q)\rangle\big\rvert_{\text{1L}} =\;\; \raisebox{-1.7 em}{\includegraphics[width=3.1cm]{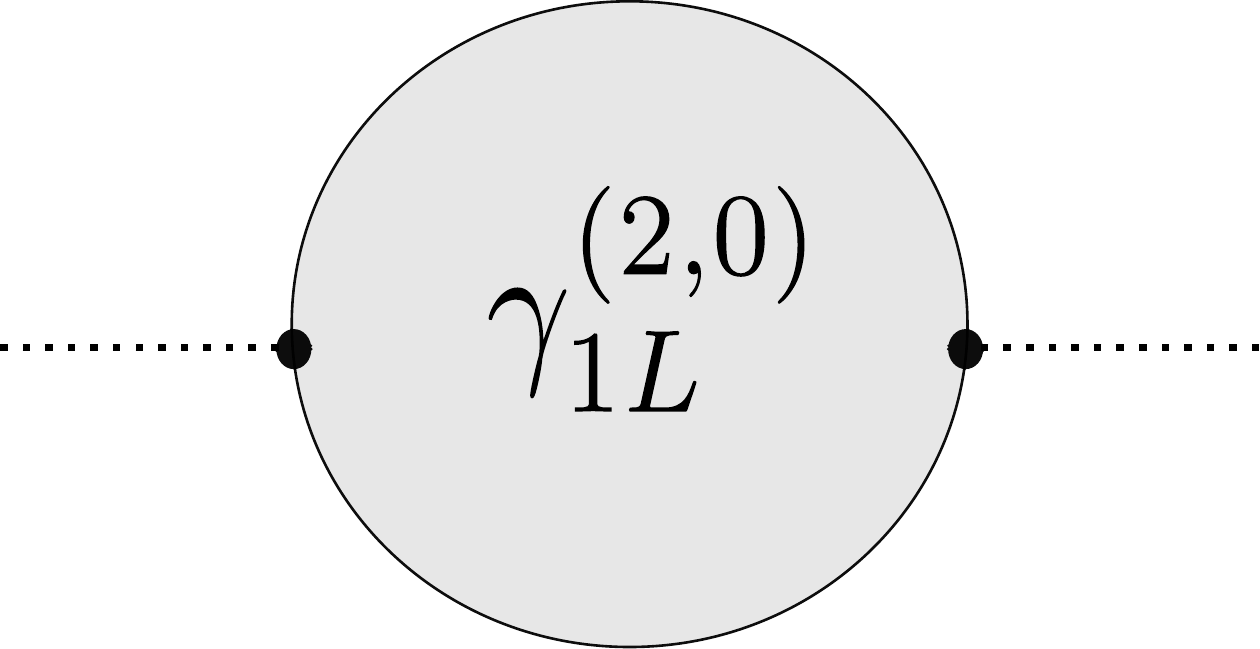}}\;\; =\;\; \raisebox{-1.7 em}{\includegraphics[width=3cm]{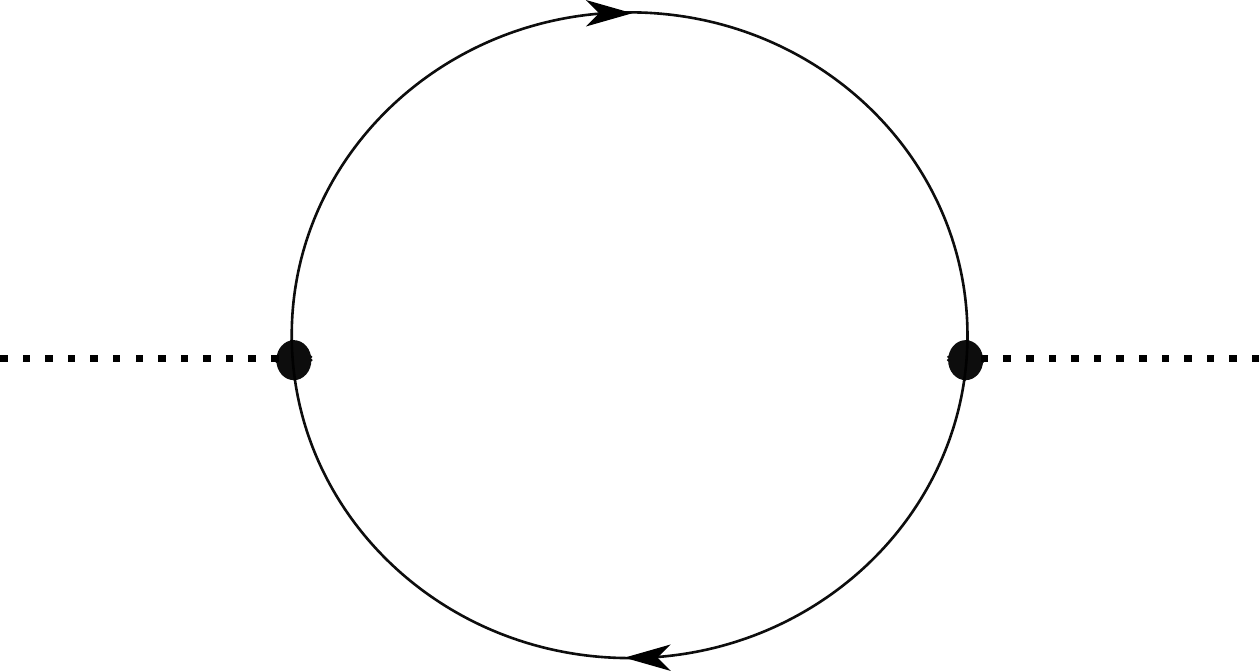}} \,.
\end{equation}
Since no internal photons appear, IR finiteness is obvious.

At two loops the entire set of 1PI diagrams can be constructed using the effective vertices $\gamma_{\mu\nu}^{(2,2)}(q_1,q_2,p_1,p_2)\big\rvert_{\text{1L}} $ and $ \gamma^{(2,0)} 
(q_1,q_2)\big\rvert_{\text{2L}}$.  Graphically they read
\begin{equation}
\begin{split}
   &\raisebox{-1.6 em}{\includegraphics[width=3cm]{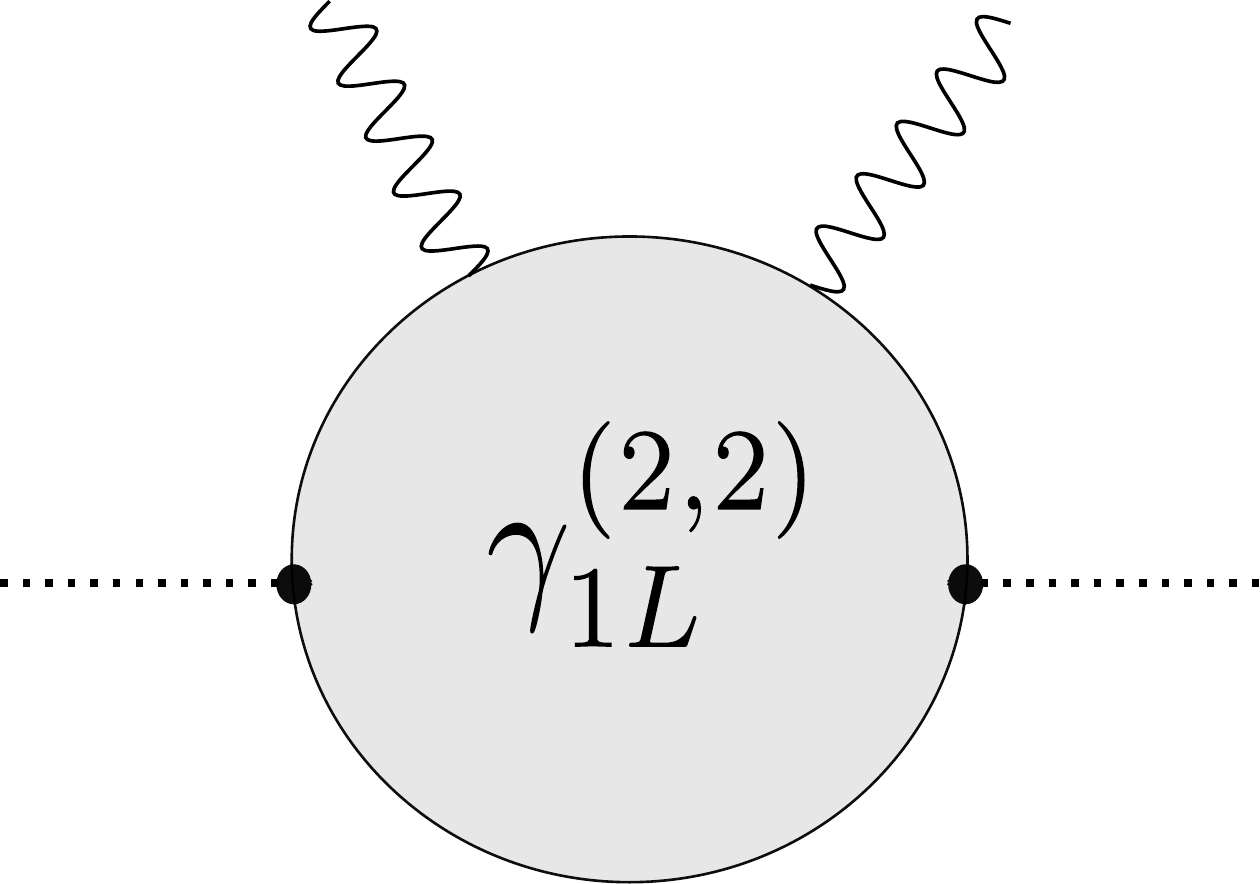}}\;\;=\;\;   \raisebox{-1.7 em}{\includegraphics[width=3cm]{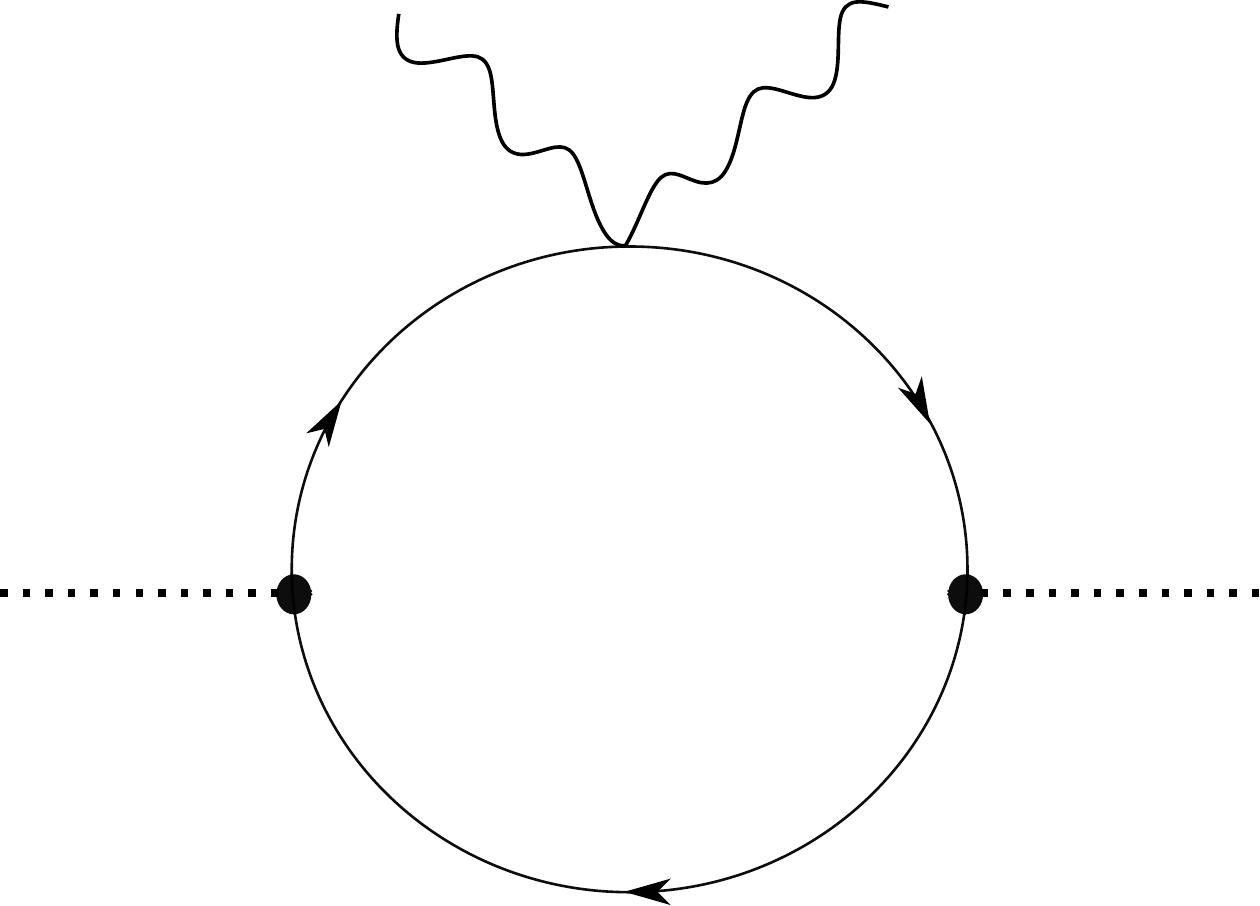}}\;\;+\;\;   \raisebox{-1.7 em}{\includegraphics[width=3cm]{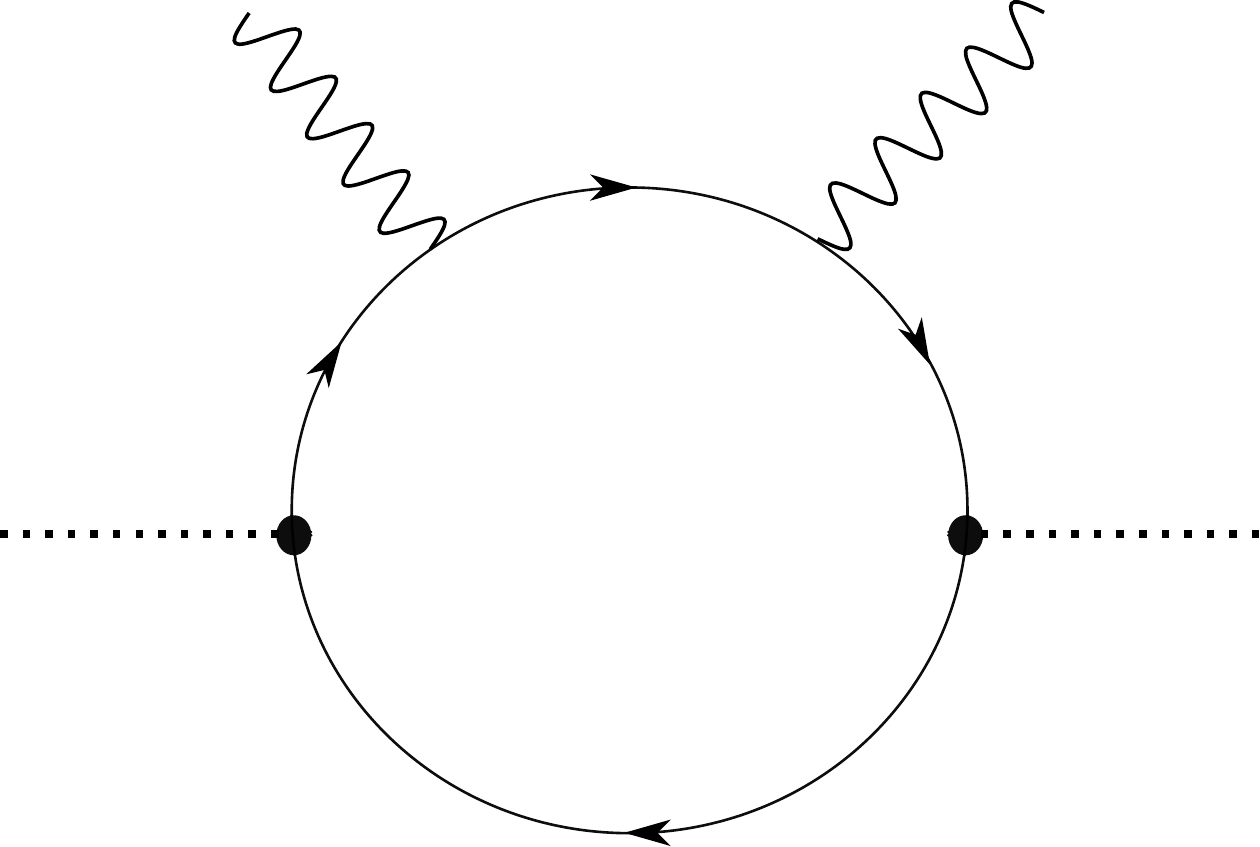}}\;\;+\;\; \raisebox{-2.9 em}{\includegraphics[width=3cm]{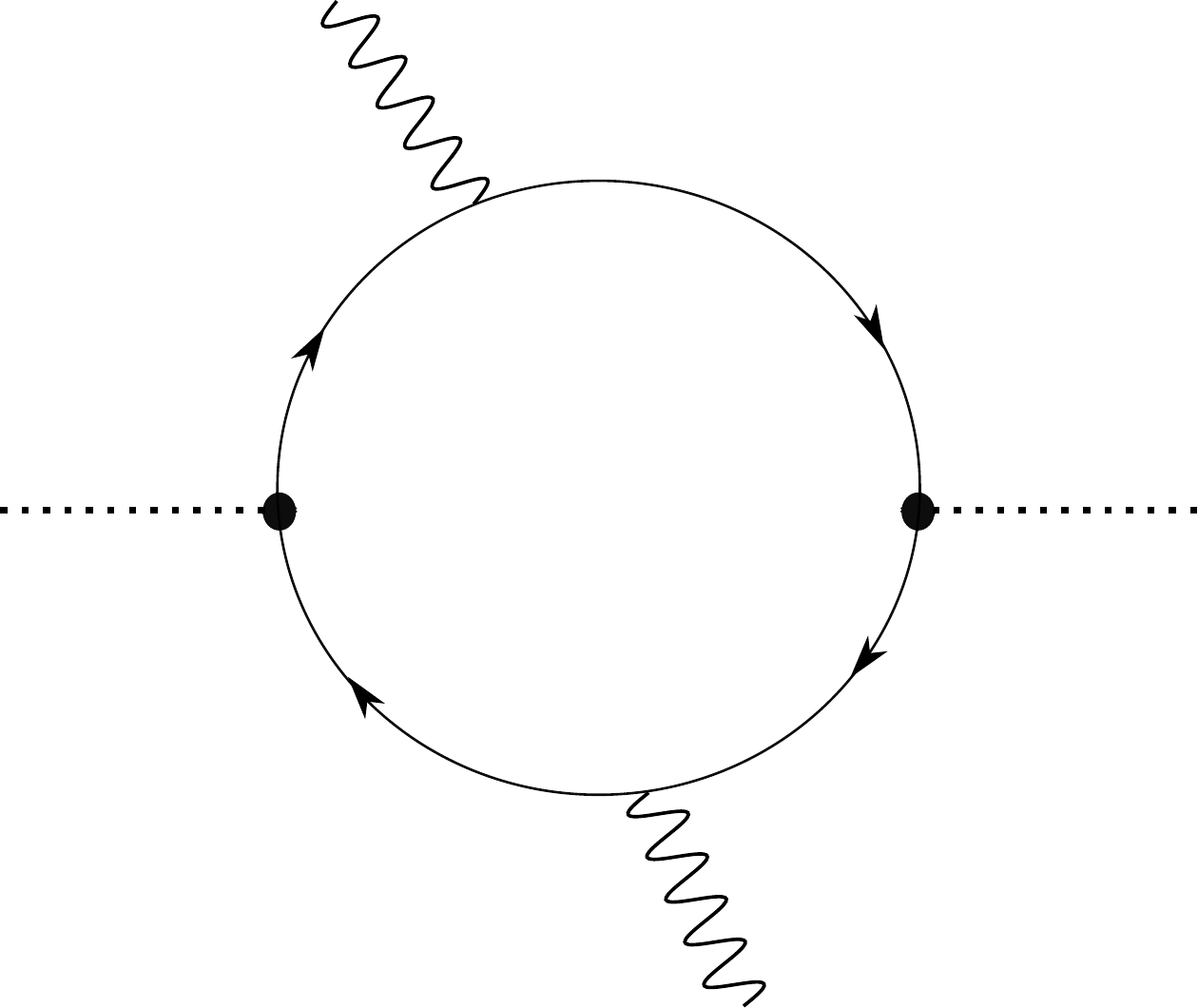}}
\end{split}
\label{}
\end{equation}
and\footnote{We have omitted to include in \eqref{OphiOphi(2L)_0} and \eqref{OphiOphi2L_1} the diagrams with the UV counter-terms for the tadpole graphs. The latter are however not necessary in dimensional regularization, where such diagrams are UV finite.}
\begin{equation}
   \raisebox{-1.6 em}{\includegraphics[width=3cm]{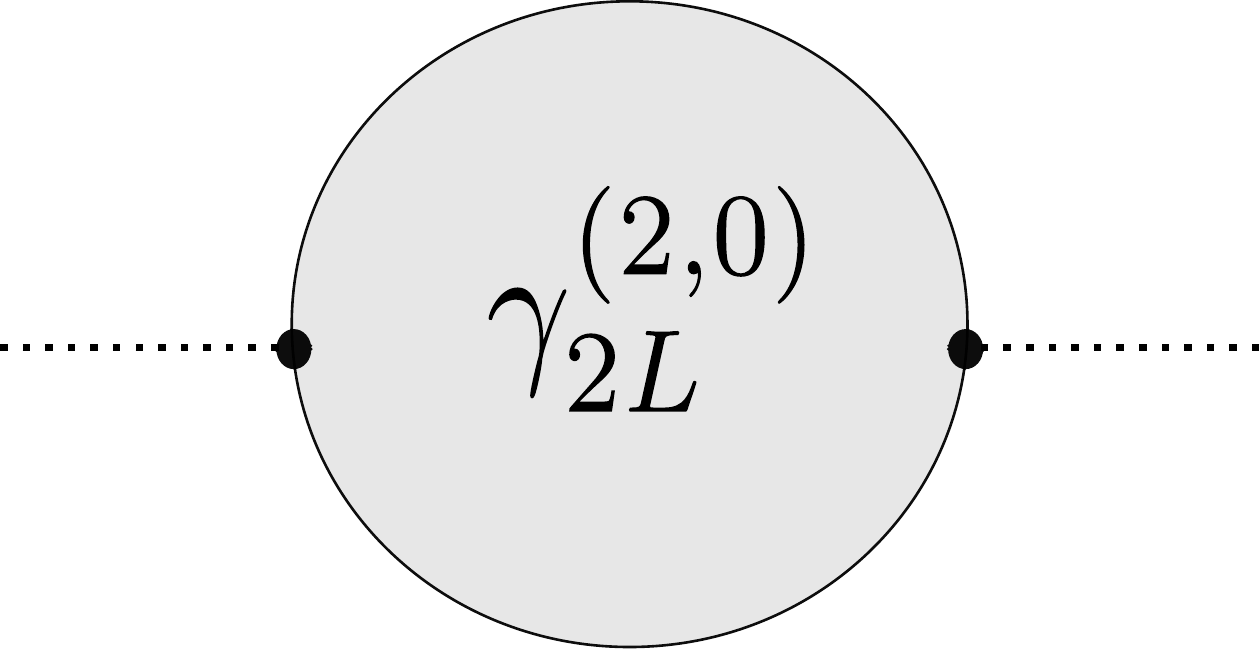}}\;\;=\; \raisebox{-1.6 em}{\includegraphics[width=4.3cm]{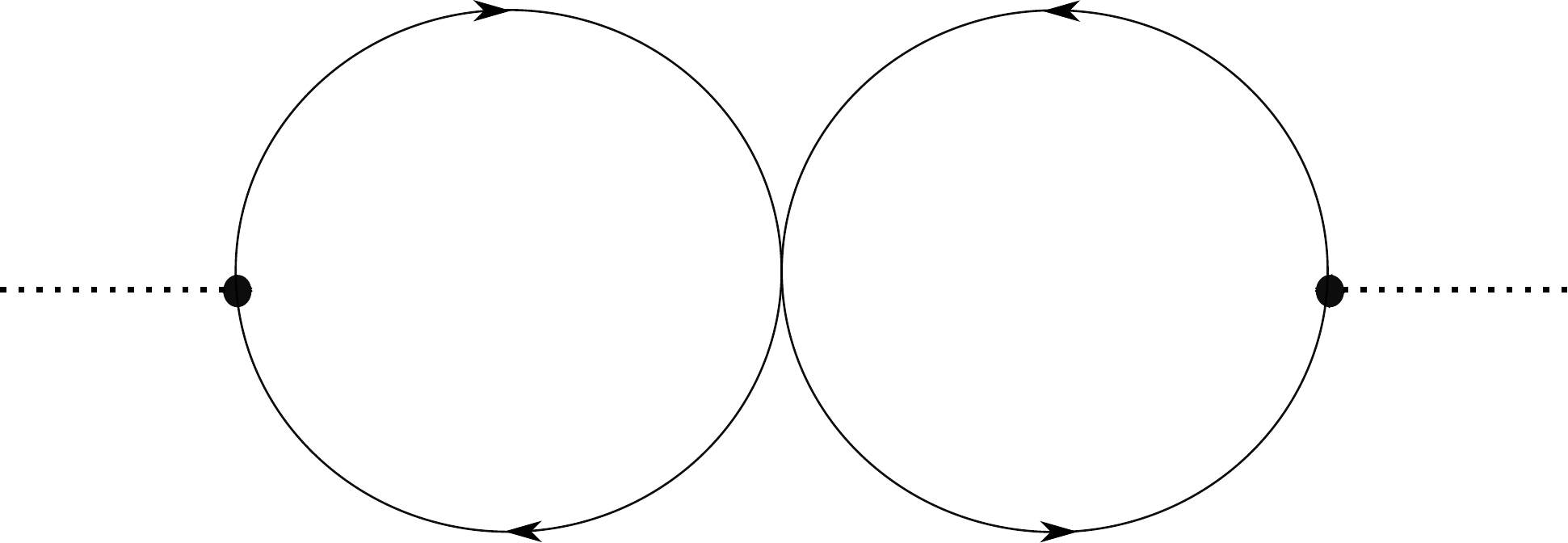}}\;\;+\;\;   \raisebox{-1.7 em}{\includegraphics[width=3cm]{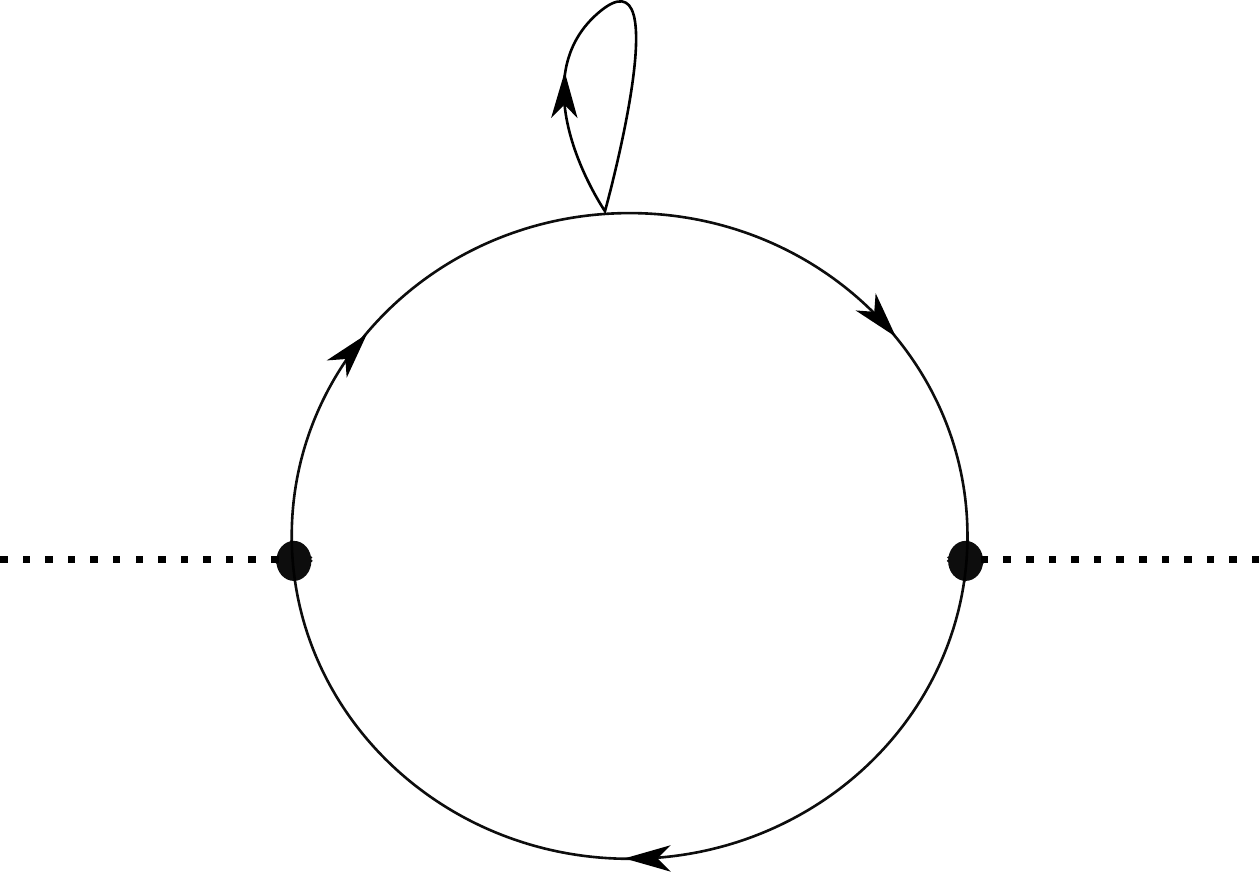}}\,.
      \label{OphiOphi(2L)_0}
\end{equation}
We have not reported the momenta flowing in the lines to avoid clutter.
By gluing together the external photon lines of $\gamma^{(2,2)}\big\rvert_{\text{1L}}$ and summing the two set of graphs, we get the two loops 1PI two-point function:
\begin{equation}
\begin{split}
&  \langle\phi^{\dagger}\phi(q)\phi^{\dagger}\phi(-q)\rangle \big\rvert_{\text{2L}}  = \;\;   \raisebox{-1.5 em}{\includegraphics[width=2.8cm]{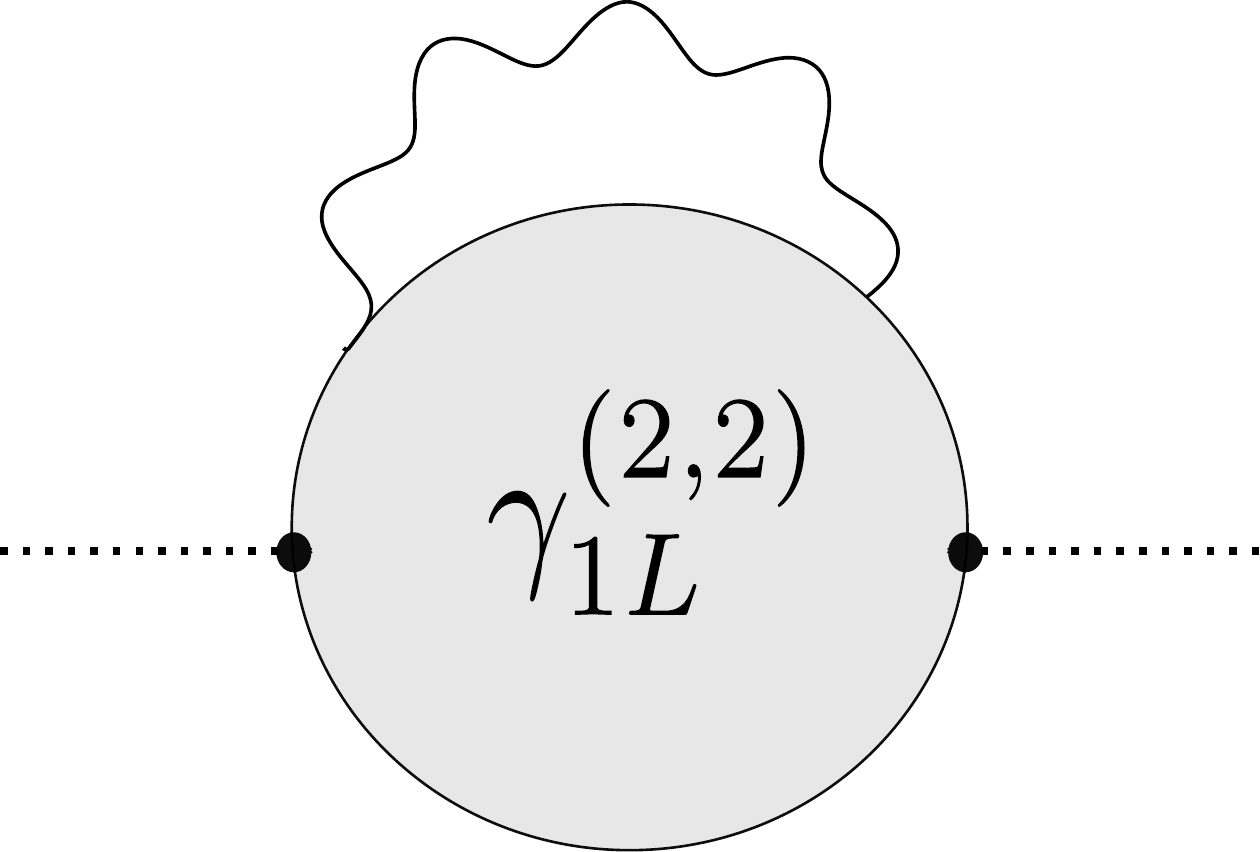}}\;\;+\;\;   \raisebox{-1.5 em}{\includegraphics[width=2.8cm]{Figs/OphiOphigamma_2.pdf}} \;\;= \\
    &  \raisebox{-1.6 em}{\includegraphics[width=2.8cm]{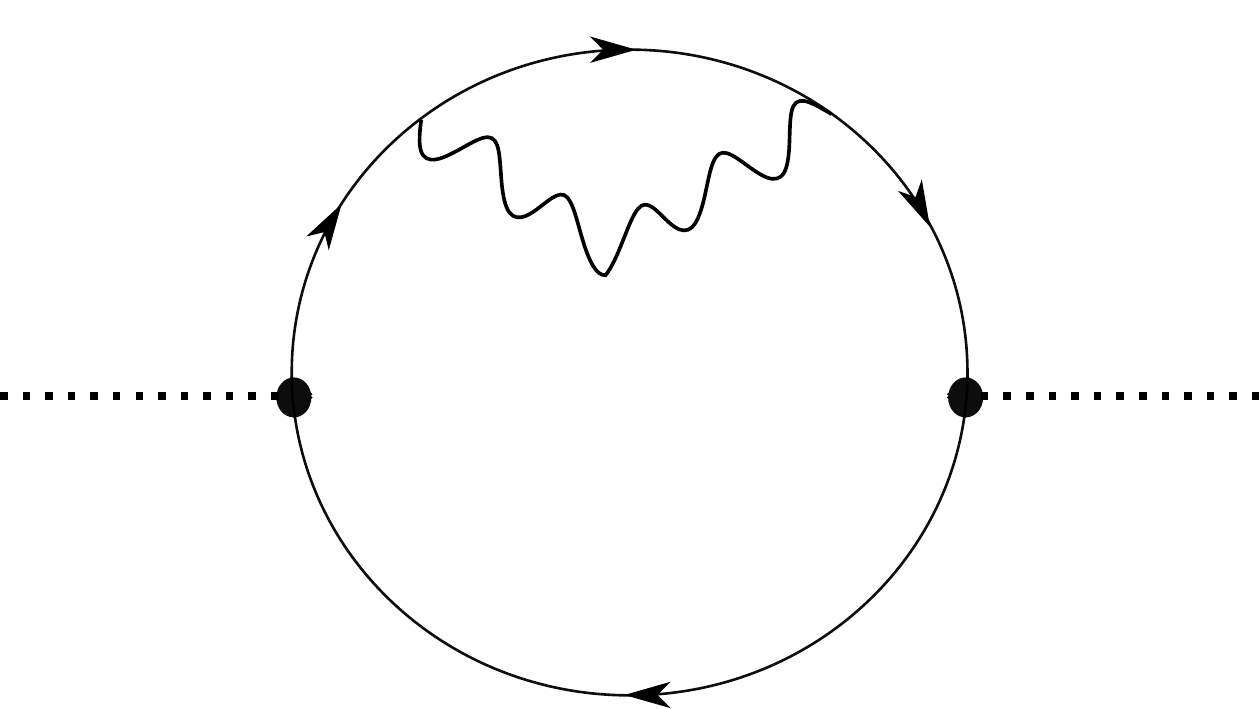}}\;\;+\;\;\raisebox{-1.6 em}{\includegraphics[width=2.8cm]{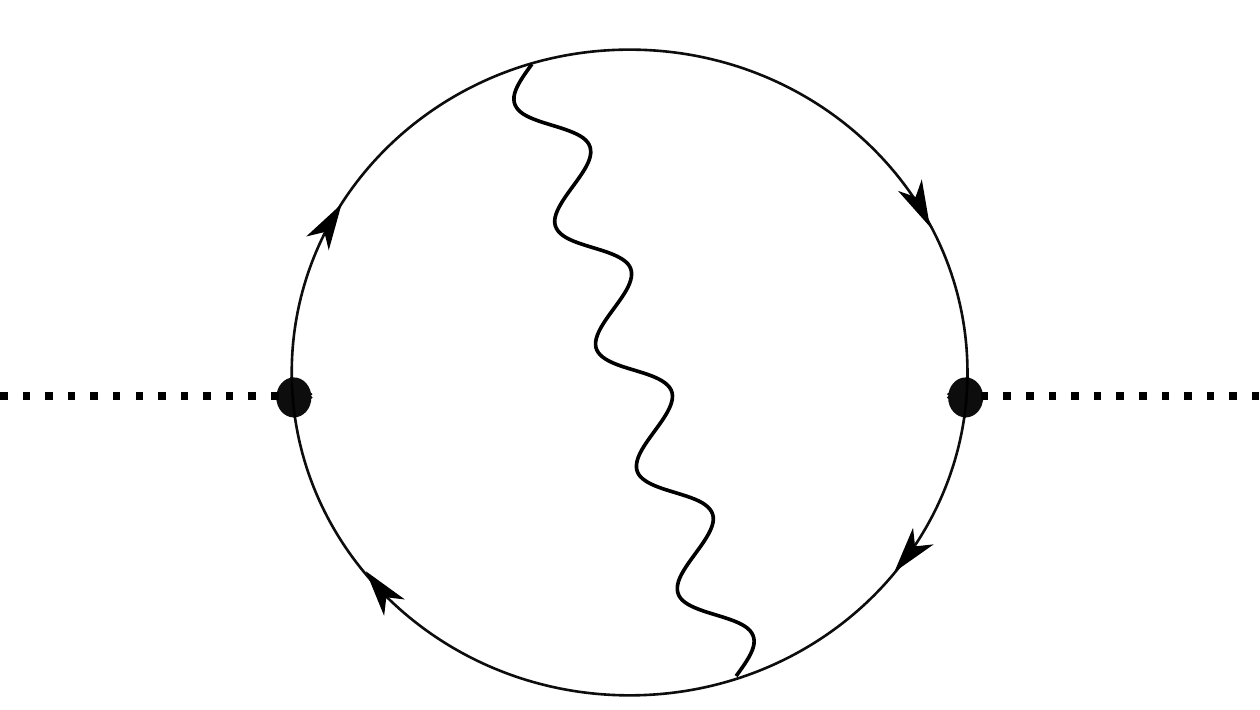}}\;\;+\;\;\raisebox{-1.6 em}{\includegraphics[width=2.8cm]{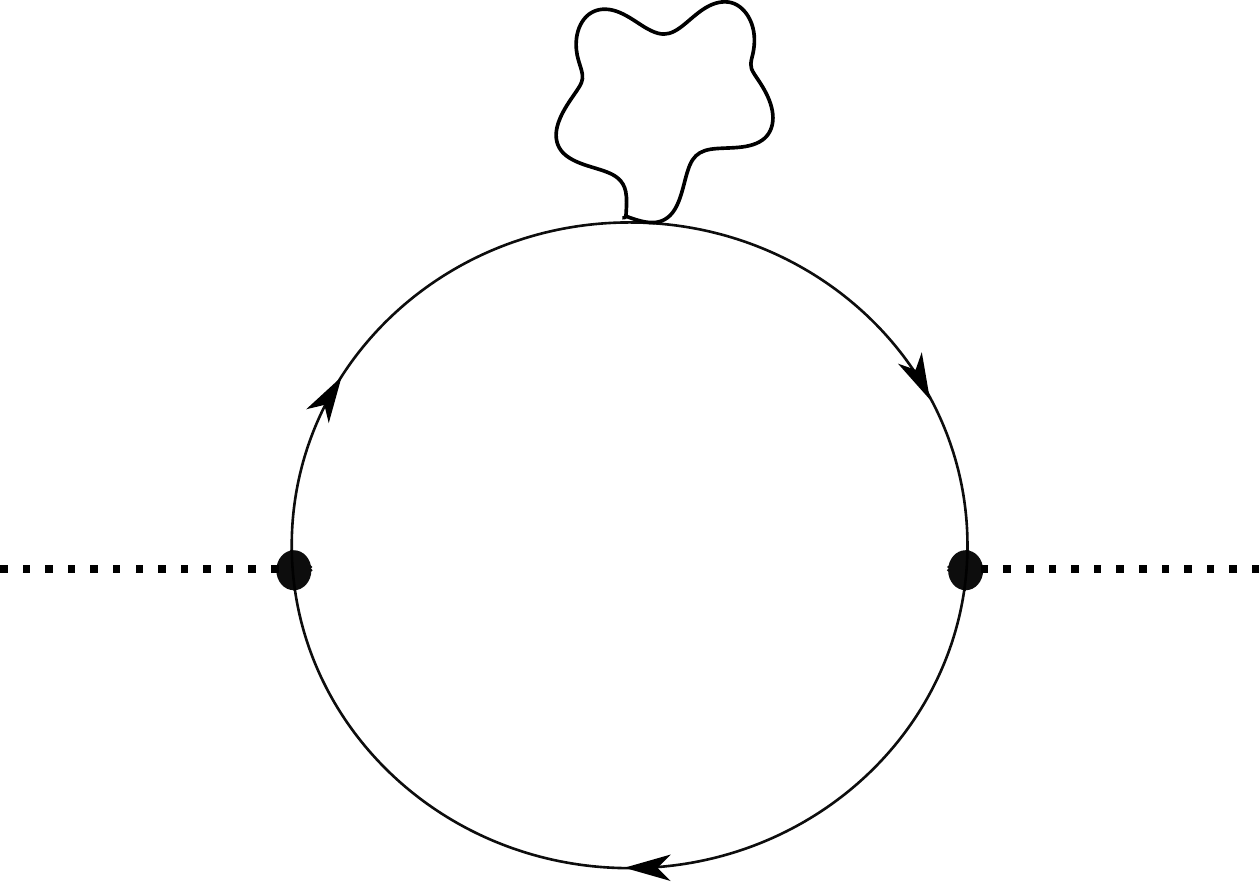}}\;\;+\\
   &\raisebox{-1.4 em}{\includegraphics[width=3.9cm]{Figs/OphiOphi1L_4_new.pdf}}\;\;+\;\;\raisebox{-1.5 em}{\includegraphics[width=2.8cm]{Figs/OphiOphi1L_5_new.pdf}}\,.
   \label{OphiOphi2L_1}
  \end{split} 
\end{equation}
Since we have at most one internal photon line in the graphs, IR finiteness is obvious. However, we can check the Ward identity \eqref{eq:gammaJi0} for the effective vertex $\gamma_{\mu\nu}^{(2,2)}\big\rvert_{\text{1L}}$. In order to prove that $\gamma_{\mu\nu}^{(2,2)}(q_1,q_2,p_1,p_2)\big\rvert_{\text{1L}}=O(p_1 p_2)$, due to the Bose symmetry in the exchange of the two external photons ($p_1\leftrightarrow p_2$), it is sufficient to check that $\gamma_{\mu\nu}^{(2,2)}\big\rvert_{\text{1L}}$ vanishes when one of the two photon momenta is zero, say $p_2=0$. 
We write $\gamma_{\mu\nu}^{(2,2)}\big\rvert_{\text{1L}} = l_{\mu\nu}^{(1)} + l_{\mu\nu}^{(2)}$, relabel $p_1\rightarrow p$, $q_1\rightarrow q$, and use momentum conservation
$q_2 = - q - p$. In this way we get\footnote{The factors $2$ and $4$ in \eqref{check_phiphiAA} represent the symmetry factors of the corresponding diagrams. These factors have been omitted before and will also be systematically omitted in the following. They have been reported here to emphasize their role in the cancellation.}
 \be
\begin{split}
    l_{\mu\nu}^{(1)}(q,-q-p,p,0) & =  2\;\;\raisebox{-1.3 em}{\includegraphics[width=2.5cm]{Figs/OphiOphiAA_2_new.pdf}}\,,\\
     l_{\mu\nu}^{(2)}(q,-q-p,p,0) & =   4\;\;\raisebox{-1.4 em}{\includegraphics[width=2.5cm]{Figs/OphiOphiAA_1_new.pdf}}\;\;+2\; \;  \raisebox{-2.4 em}{\includegraphics[width=2.5cm]{Figs/OphiOphiAA_3_new.pdf}} \,.
\end{split}
  \label{check_phiphiAA}
   \ee
After standard manipulations it is straightforward to find that
\be
\begin{split}
 l_{\mu\nu}^{(2)}(q,-q-p,p,0) & = -2e^2\int\frac{d^3k}{(2\pi)^3}\frac{
  -2g^{\mu\nu}}{(k^2+m^2)((k-p)^2+m^2)((k-p-q)^2+m^2)} \\
  & =  -  l_{\mu\nu}^{(1)}(q,-q-p,p,0)\,,
\end{split}
 \ee 
 proving in this way that $\gamma_{\mu\nu}^{(2,2)}(q,-q-p,p,0)\big\rvert_{\text{1L}} =0$ for any value of $p$ and $q$.

At three loops we get IR divergent diagrams which sum to a finite result. The vertices entering at this order are $\gamma^{(2,0)}\big\rvert_{\text{3L}}$, $\gamma_{\mu_1\mu_2}^{(2,2)}\big\rvert_{\text{2L}}$, $\gamma_{\mu_1\cdots\mu_4}^{(2,4)}\big\rvert_{\text{1L}}$, $\gamma_{\mu\nu}^{(1,2)}\big\rvert_{\text{1L}}$, $\gamma_{\mu\nu}^{(2,2)}\big\rvert_{\text{1L}}$ and $\gamma_{\mu\nu}^{(0,2)}\big\rvert_{\text{1L}}$.
We do not write all of them explicitly, but it is easy to check that they produce, when combined together, the 1PI part of the correlator as follows:
\begin{equation}
	\begin{split}
		\langle \phi^{\dagger}\phi(q)\phi^{\dagger}\phi(-q)\rangle\big\rvert_{\text{3L}}  = & \;\;\mathop{\raisebox{-1.5 em}{\includegraphics[width=2.8cm]{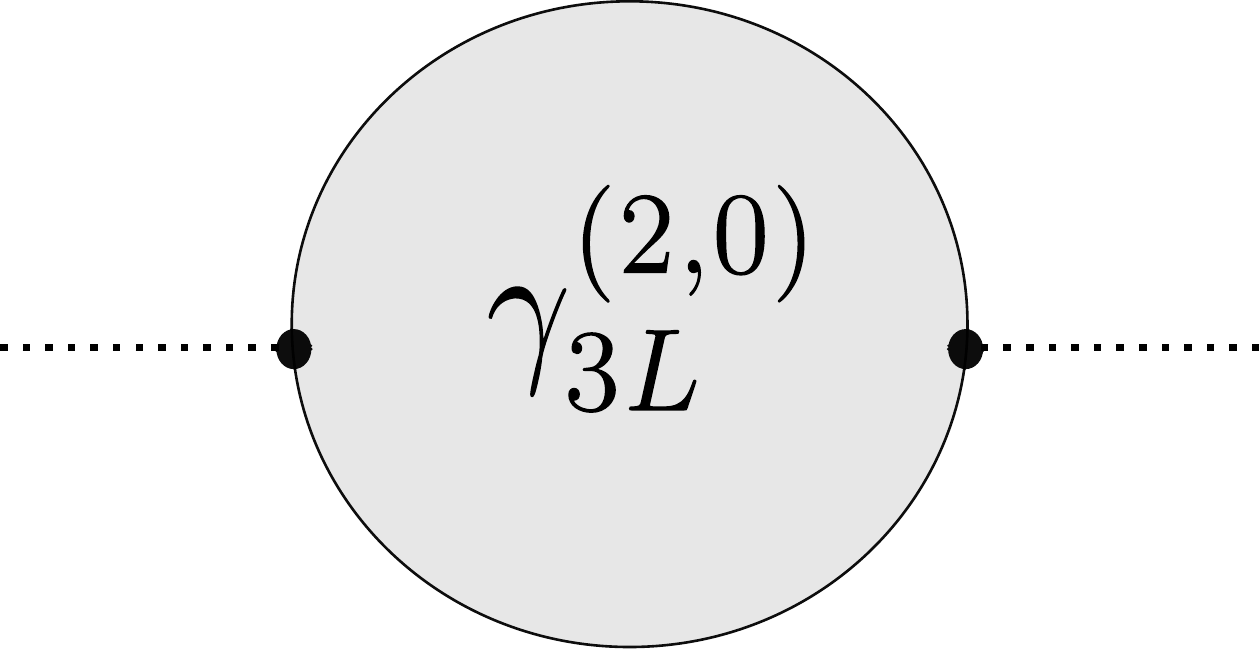}} }_{(a)}\;\;+\;\mathop{\raisebox{-1.5 em}{\includegraphics[width=2.8cm]{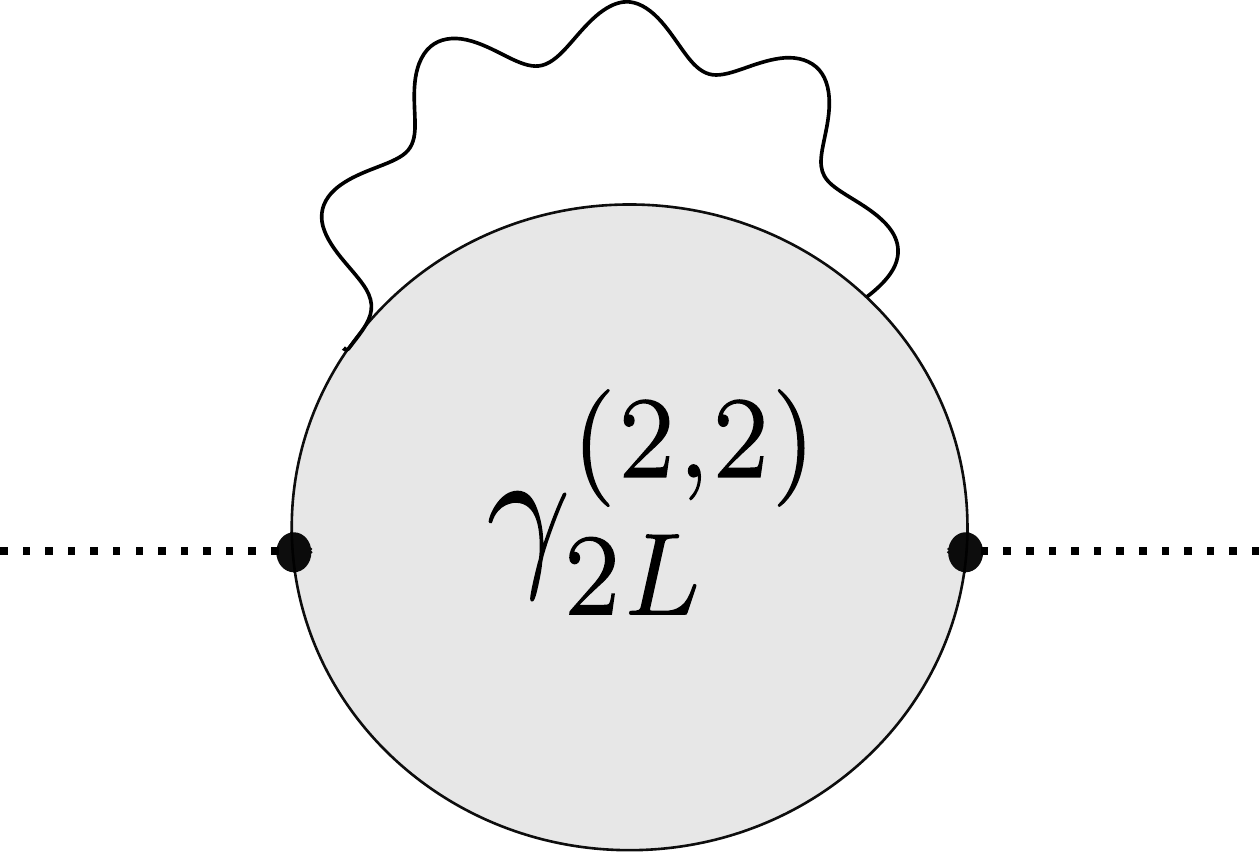}}}_{(b)}\;\;+\; \mathop{\raisebox{-2.4em}{\includegraphics[width=2.8cm]{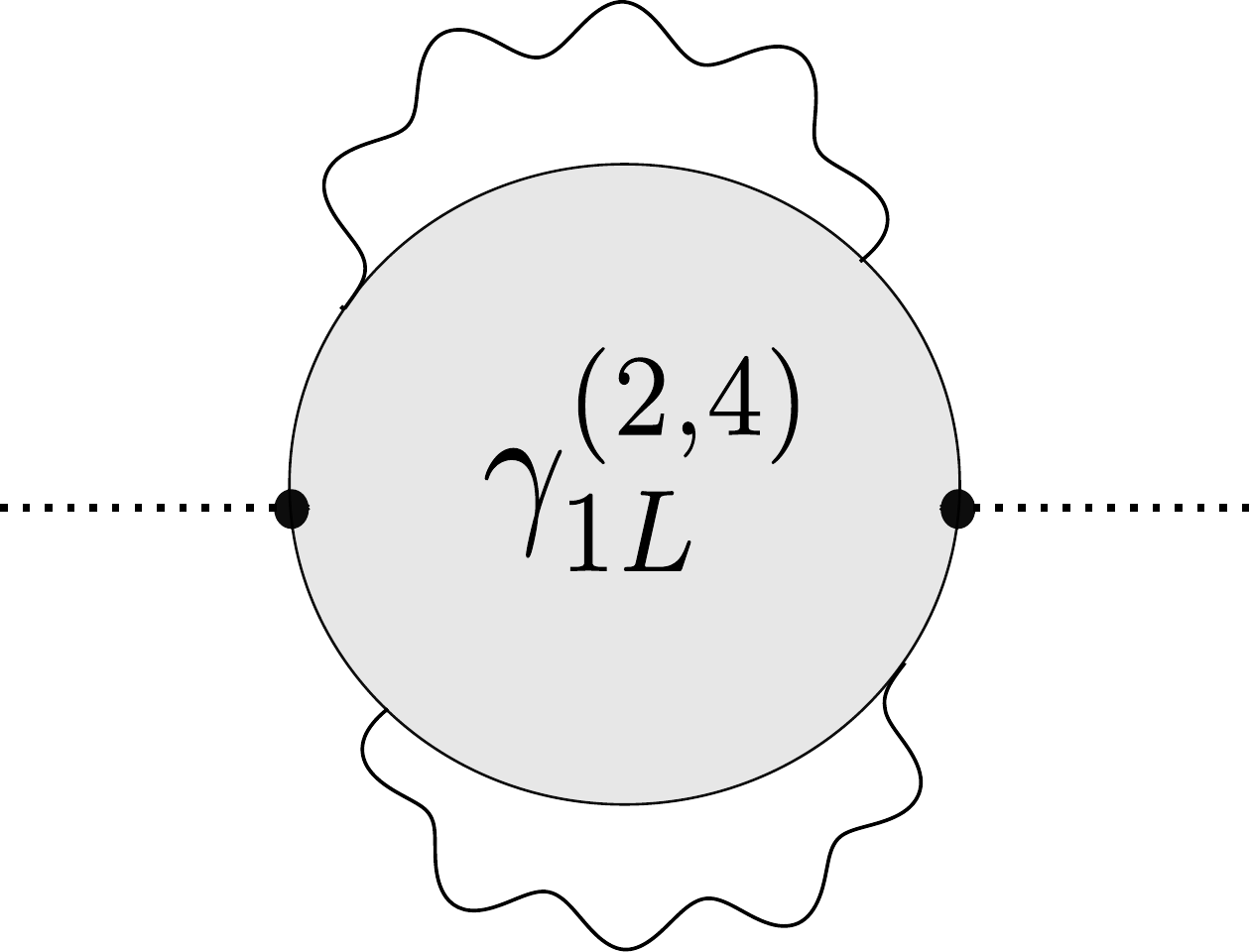}}}_{(c)}\;+\\
		&\;\mathop{\raisebox{-1.5 em}{\includegraphics[width=2.8cm]{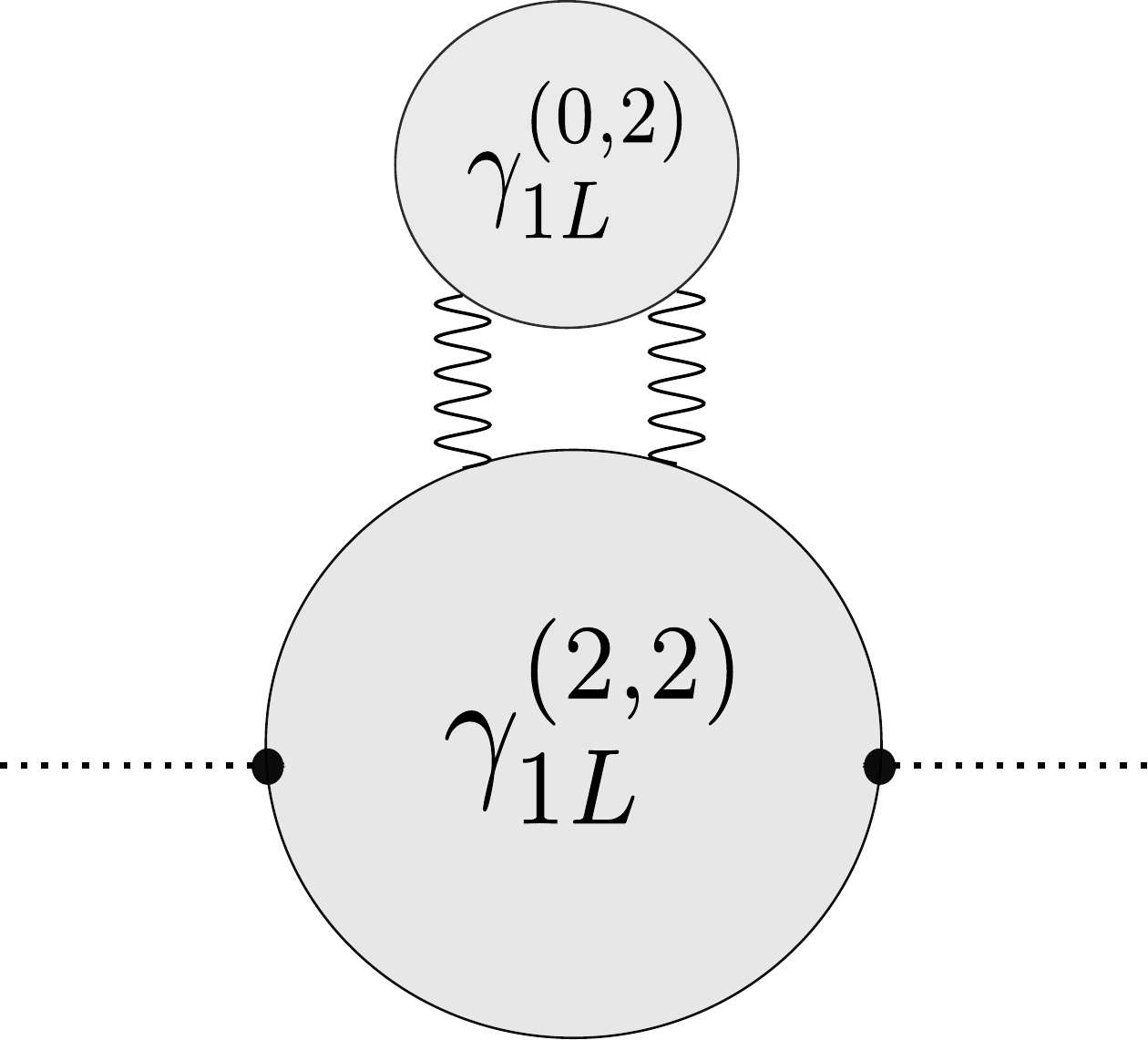}}}_{(d)}\;\;+\;\;\mathop{\raisebox{-1.4 em}{\includegraphics[width=5.8cm]{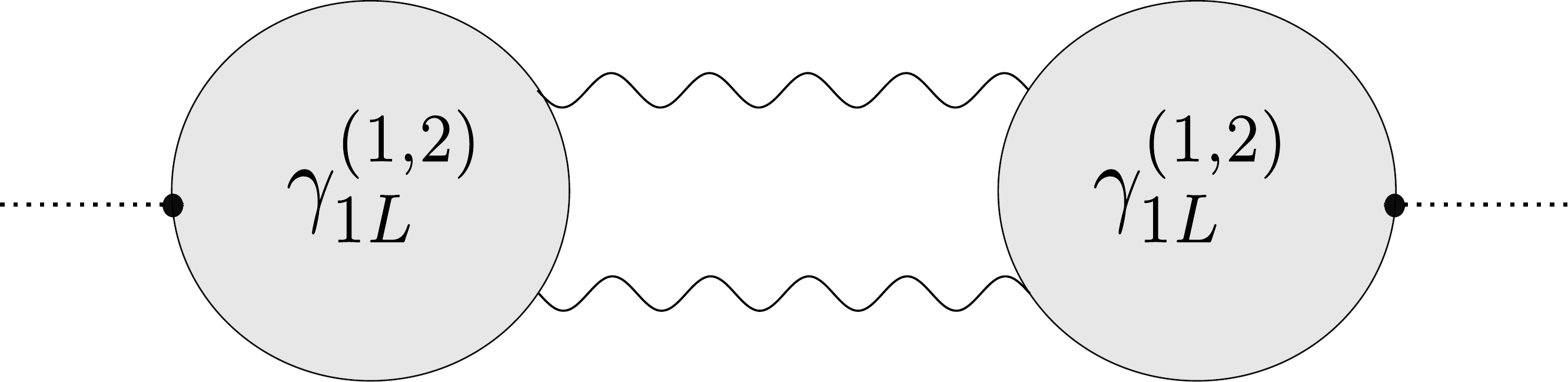}}}_{(e)}\,.
	\end{split}
\label{OO3L}
\end{equation}
Let us briefly discuss the IR properties of these diagrams, without listing them one by one. The groups $(a)$ and $(b)$  are composed by diagrams with 0 and 1  internal photon propagators, respectively, and are hence trivially IR finite. The group $(c)$ is composed by three loop diagrams with $2$ internal photon lines with independent virtual momenta, which 
are then individually IR finite. The groups of diagrams (d) and (e) are the ones where IR divergent Feynman diagrams appear. In group $(d)$ they arise  for every choice of the external momentum $q$. The cancellation of IR divergences is evident from the IR behaviour of $\gamma_{1L}^{(2,2)}$ just shown. Furthermore, $\gamma^{(0,2)}_{1L}$ coincides with the one-loop photon $2$-point function:
\be
 \raisebox{-1.7 em}{\includegraphics[width=4.2cm]{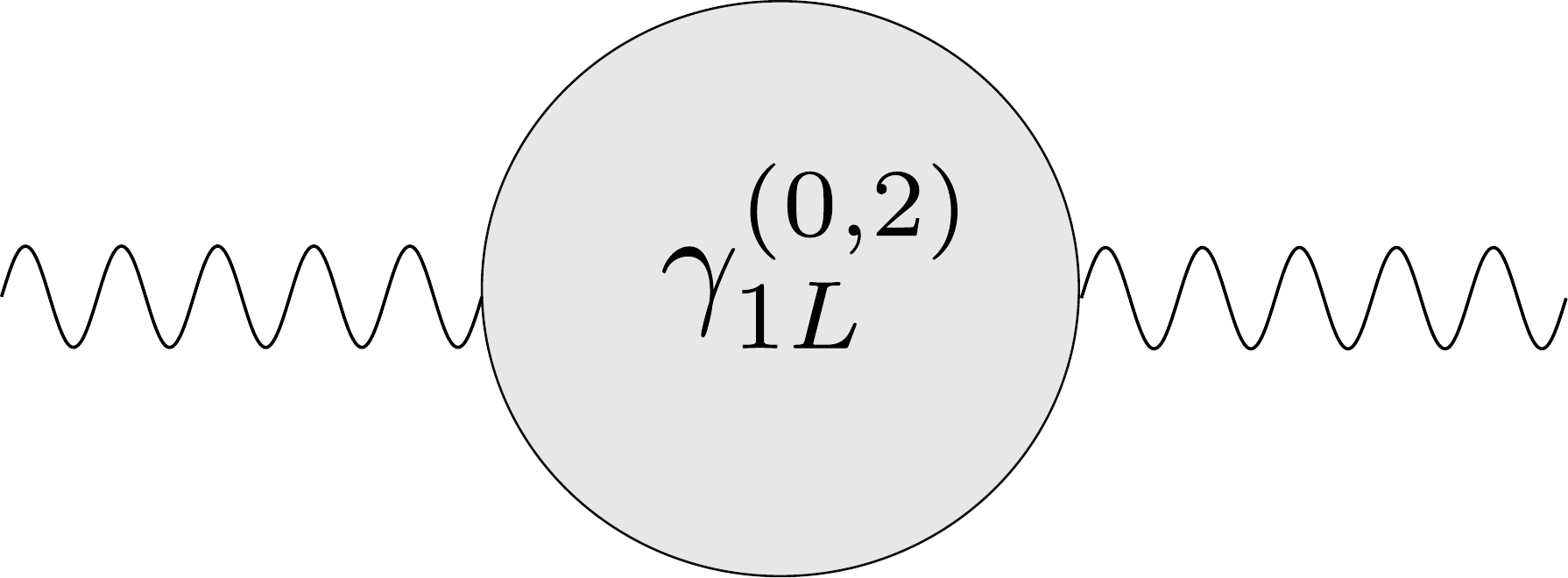}}\;\;=\; \raisebox{-1.8 em}{\includegraphics[width=4.2cm]{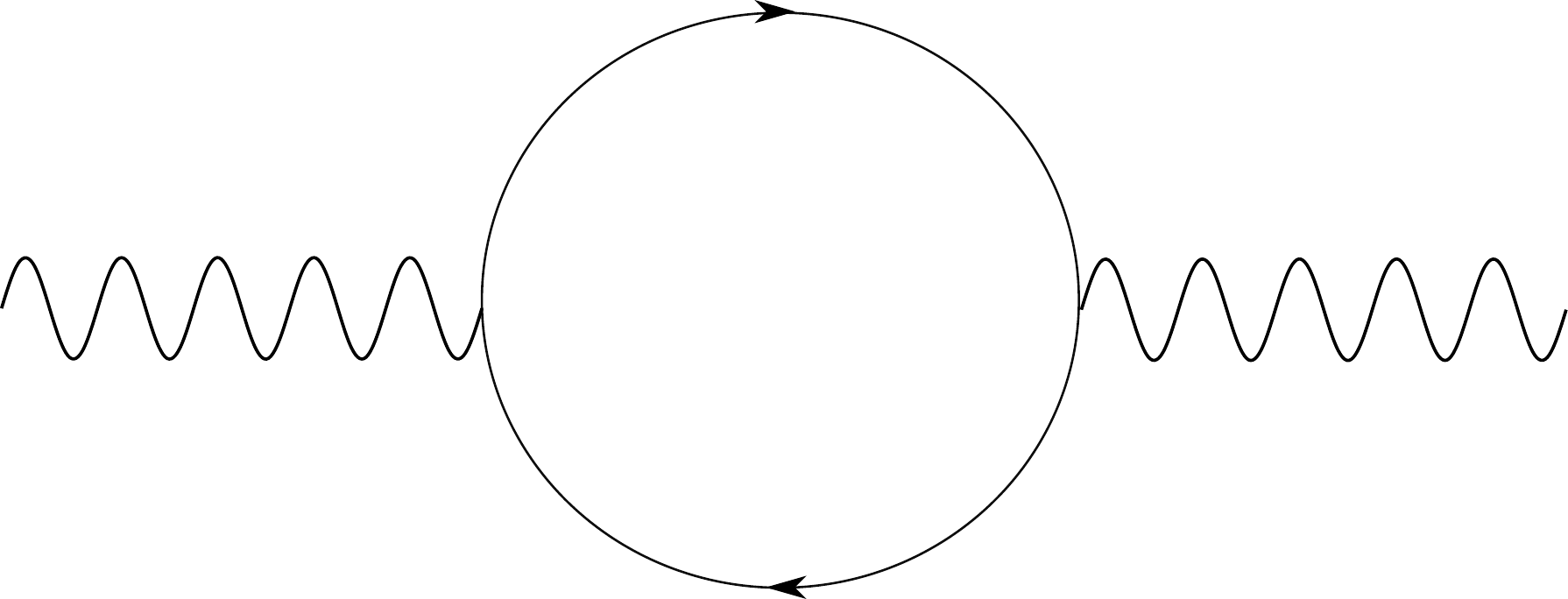}}\;\;+\;\;   \raisebox{-0.15 em}{\includegraphics[width=4.2cm]{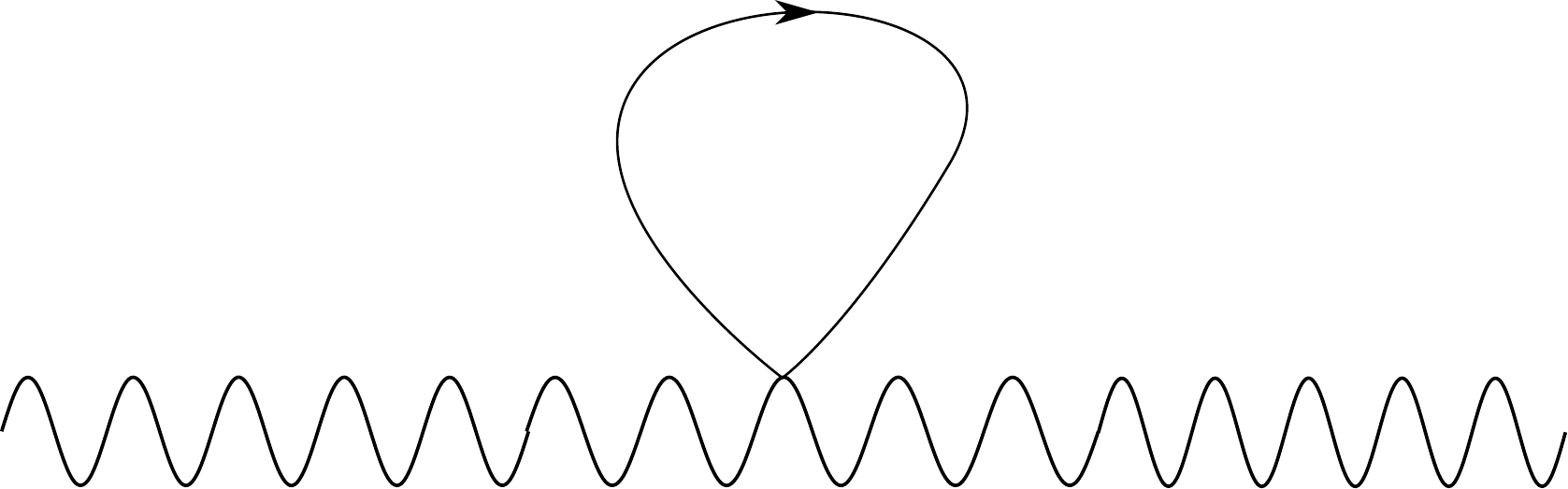}} \,,
\ee
and two more powers of $p$ arise from the transversality of the photon propagator. IR divergences in single Feynman diagrams appear in group $(e)$  when the external momentum $q$ vanishes. Their sum is guaranteed to vanish thanks to the powers of $p_i$ coming form the two $\gamma^{(1,2)}$, as predicted by \eqref{eq:gammaJi0}. We can check that relation by expanding $\gamma_{\mu\nu}^{(1,2)}(-p_1-p_2,p_1,p_2)\big\rvert_{\text{1L}}$ for small $p_1$ and $p_2$ and see that the resulting function is $O(p_1p_2)$. Indeed, after standard manipulations, we get
\beaa
  &&  \gamma_{\mu\nu}^{(1,2)}(-p_1-p_2,p_1,p_2)\big\rvert_{\text{1L}}= \; \raisebox{-2.3 em}{\includegraphics[width=3cm]{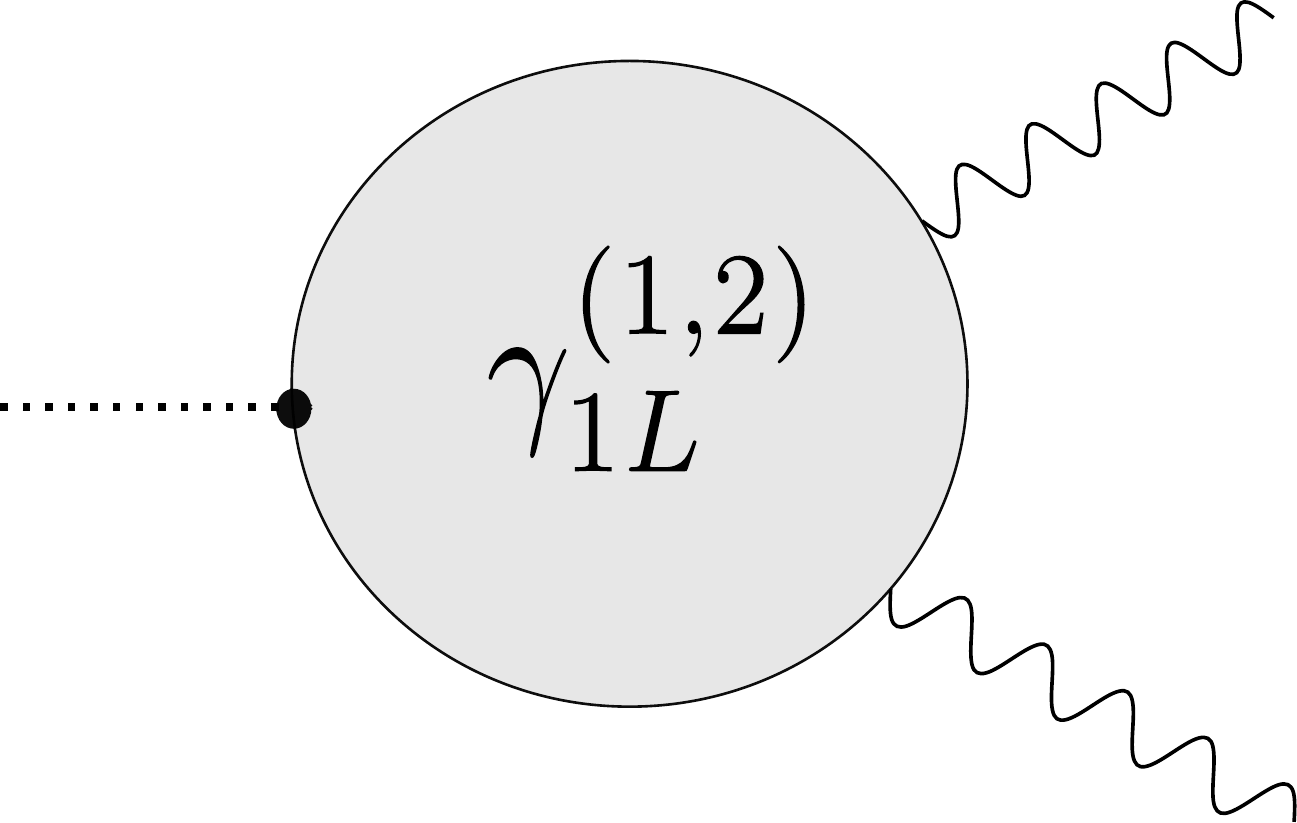}} \!\! = \;\raisebox{-2.3 em}{\includegraphics[width=3cm]{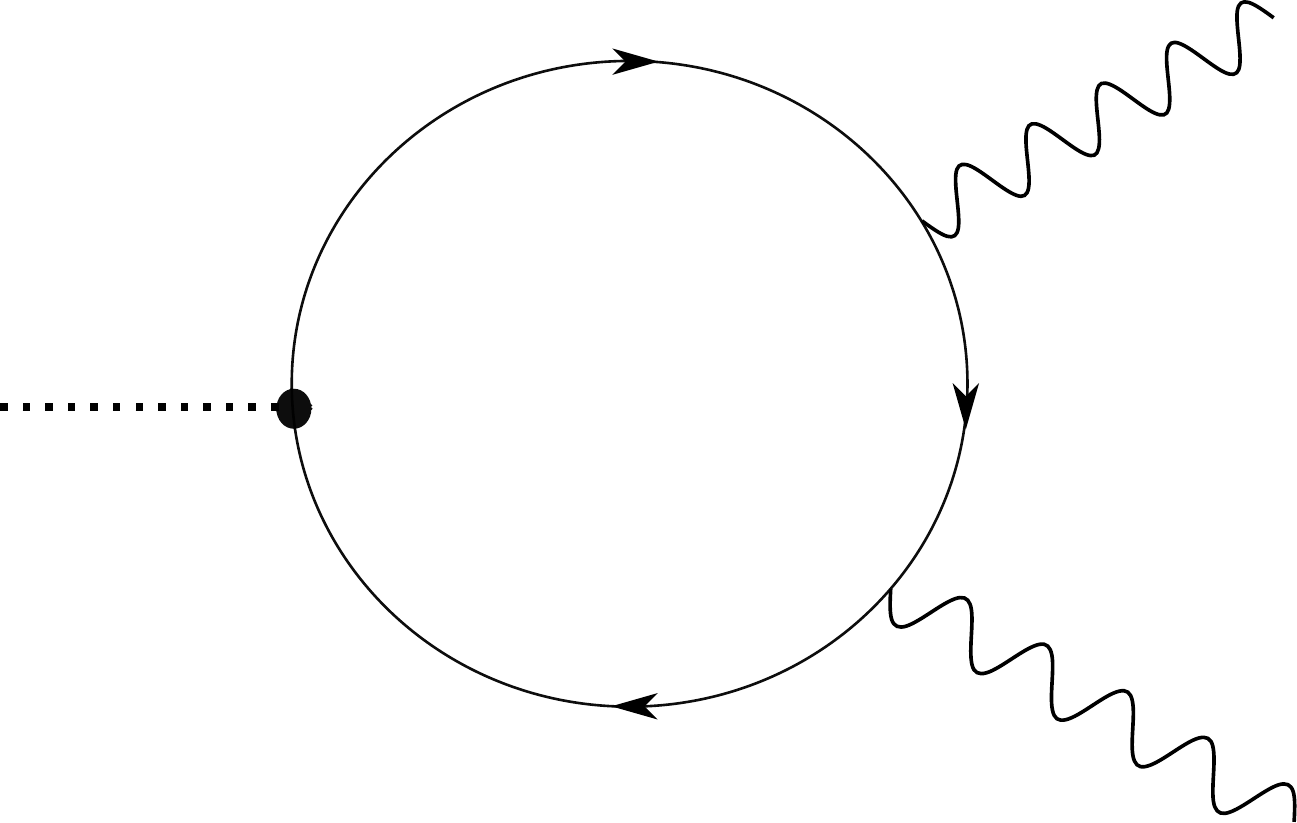}} \!\! + \; \raisebox{-1.6 em}{\includegraphics[width=3cm]{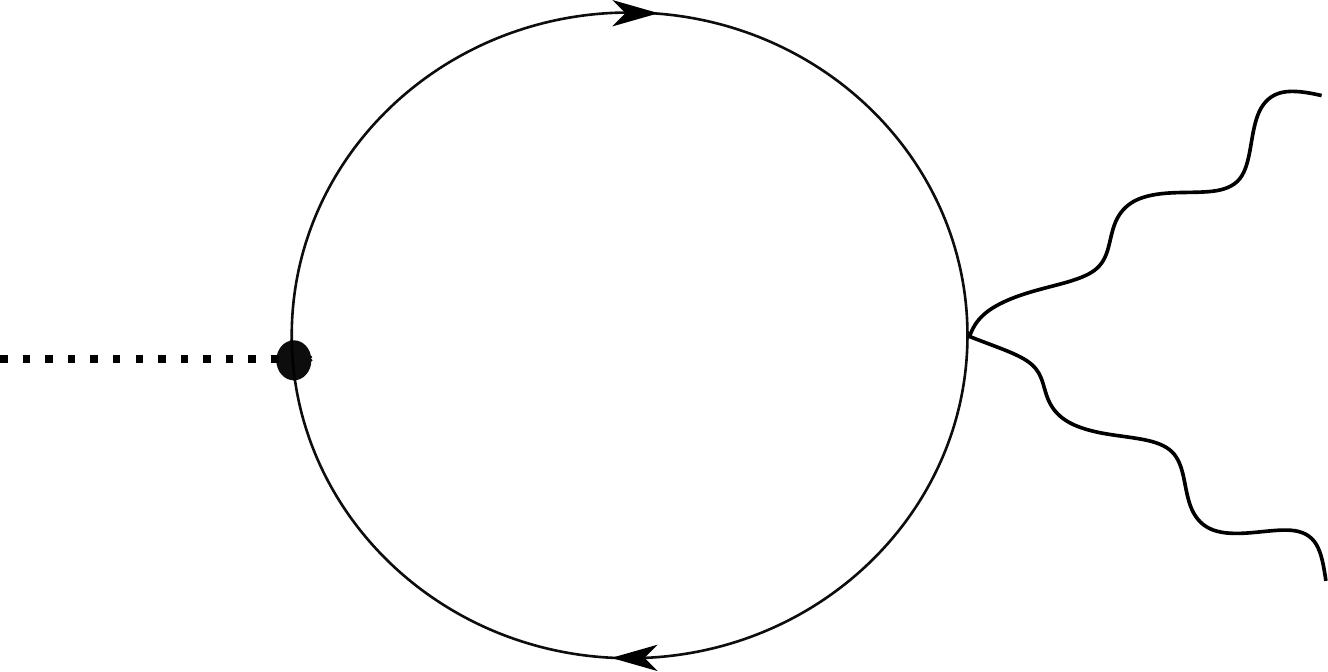}} \nn  \\ [3mm]
        & & =  \int\frac{d^3k}{(2\pi)^3}\frac{e^2}{(k^2+m^2)((k-p_1-p_2)^2+m^2)}\left(\frac{(2k-p_1)^{\mu}(2k-2p_1-p_2)^{\nu}+(\mu\leftrightarrow\nu)}{((k-p_1)^2+m^2)}-2g^{\mu\nu}\right) \nn \\ [3mm]
      && =\frac{e^2}{96\pi m^{3}}(2g^{\mu\nu}p_1\cdot p_2 - p_1^{\mu}p_2^{\nu}-p_1^{\nu}p_2^{\mu})+O(p_1^2 p_2, p_1 p_2^2)\;\;,
         \label{check_phiAA}
 \eeaa
 completing the explicit check of the cancellation of IR divergences up to three-loop order for the correlator $\langle \phi^{\dagger}\phi(q)\phi^{\dagger}\phi(-q)\rangle$.

\subsection{$\langle F_{\mu\nu}\phi^{\dagger}\phi(q)F_{\rho\sigma}\phi^{\dagger}\phi(-q) \rangle$}
\label{subsec:phiphiF}

As a second example we consider a gauge-invariant tensor operator composed of both elementary matter and gauge field operators. 
At each order in perturbation theory, all the 1PI diagrams entering the correlator can be written as sum of effective vertices $\gamma^{(k,n)}$, with $k=0,1,2$, evaluated at 
the appropriate order. We will not show such decomposition, which can be obtained by properly rearranging the Feynman diagrams.
We instead focus on a given subset $\gamma^{(k,n)}$ and explicitly show the validity of \eqref{eq:gammaJi0}. 
It should now be clear that potentially IR divergent graphs are obtained when two $\gamma^{(k,n)}$ are connected by at least two photon lines.
Consider for instance the diagrams obtained by gluing two vertices of the kind $\gamma^{(1,n)}$ with $n\geq 2$ photon lines. 
By Lorentz invariance and charge conjugation symmetry $\gamma^{(1,2)}=0$ to all orders, so let us consider $n=3$. 
The vertex $\gamma^{(1,3)}$ is non-trivial starting from one-loop level, so we can restrict to this order.\footnote{In general the operator $F_{\mu \nu}\phi^\dagger \phi$ can mix with $F_{\mu\nu}$ already at one-loop level, but such one-loop mixing is absent in dimensional regularization.}
 By gluing together a pair of two vertices $\gamma_{\mu\nu\alpha_1\alpha_2\alpha_3}^{(1,3)}(q_1,p_1,p_2,p_3)\big\rvert_{\text{1L}}$ 
we get diagrams of the form
\begin{equation}
    \raisebox{-1.3em}{\includegraphics[width=8.5cm]{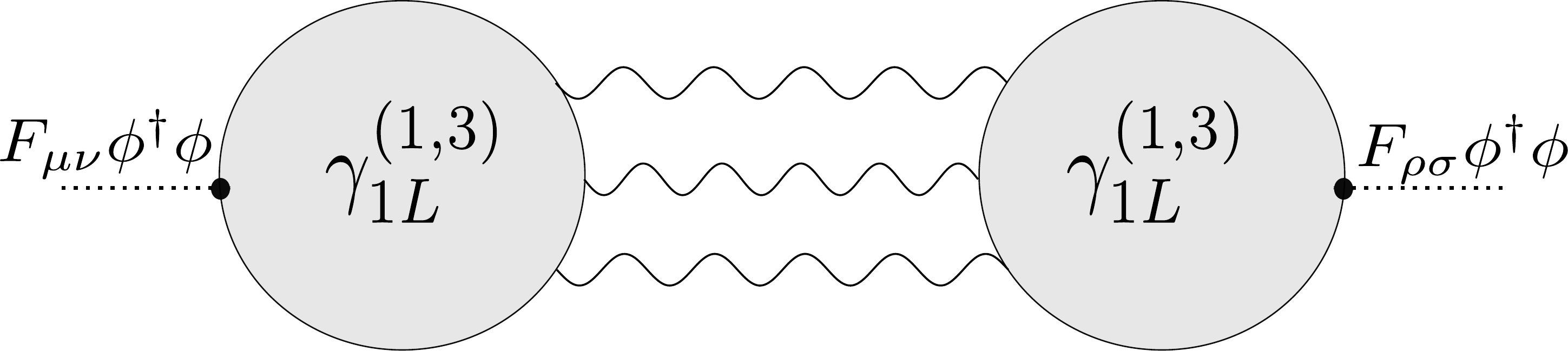}}.
    \label{gluing4}
\end{equation}
As before, the individual diagrams entering \eqref{gluing4} are IR divergent, but their cancellation is guaranteed if we verify \eqref{eq:gammaJi0} for $\gamma^{(1,3)}$. 
For simplicity let us set $q_1=0$, which is the worst case scenario as far as IR divergences are concerned. We then get
 \begin{equation}
 \begin{split}
   & \gamma_{\mu\nu\alpha_1\alpha_2\alpha_3}^{(1,3)}(0,p_1,p_2,p_3=-p_1-p_2)=\;\;\raisebox{-2.05 em}{\includegraphics[width=3cm]{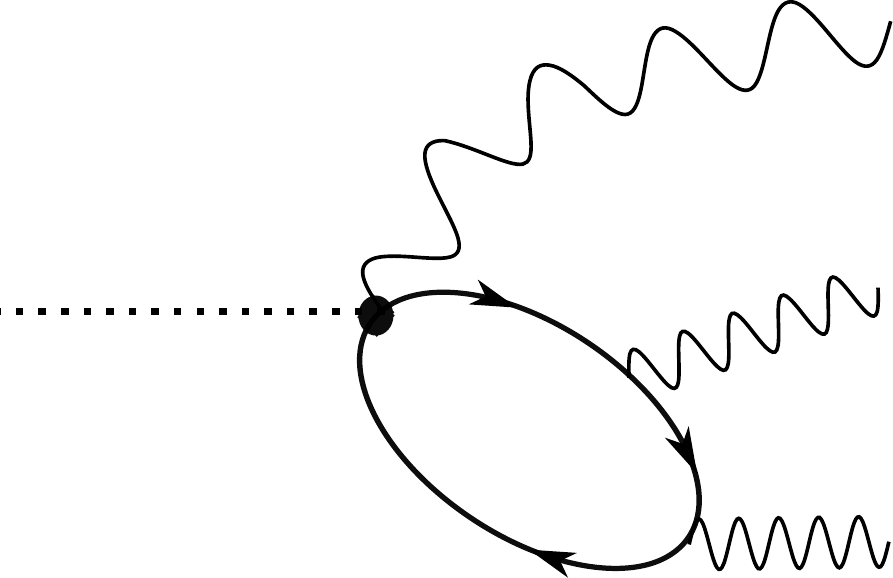}} + \raisebox{-2.25em}{\includegraphics[width=3cm]{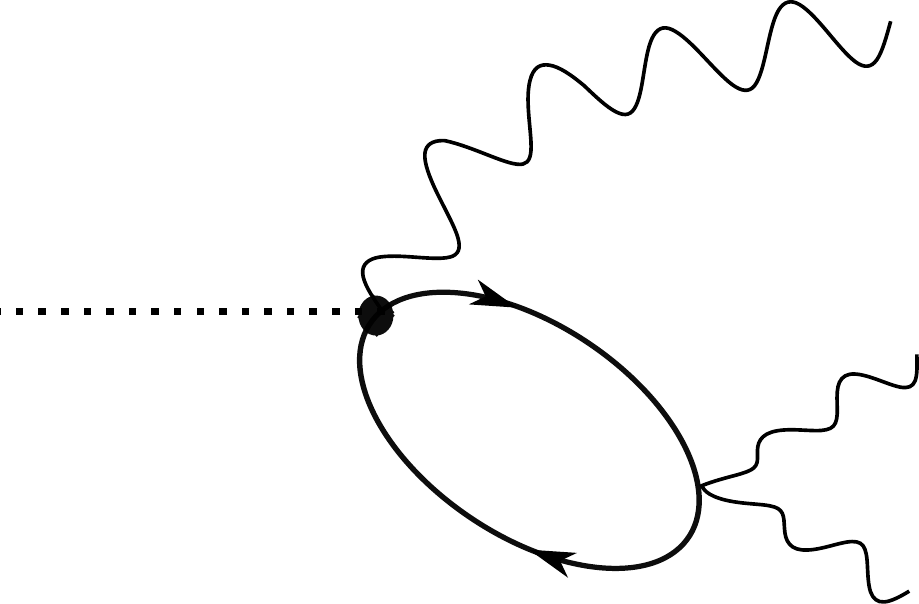}}\\[2mm]
   &= (p_{1\mu}g_{\nu\alpha_1}-p_{1\nu}g_{\mu\alpha_1})  \gamma_{\alpha_2\alpha_3}^{(\phi^\dagger \phi,2)}(-p_2-p_3,p_2,p_3) +(p_1\leftrightarrow p_2)+(p_1\leftrightarrow p_3)\,,
      \end{split} 
      \label{check_fphiphi}
\end{equation}
where $\gamma_{\alpha_2\alpha_3}^{(\phi^\dagger \phi,2)}$ is precisely the effective vertex  
defined in \eqref{check_phiAA}.\footnote{It is here denoted $\gamma_{\alpha_2\alpha_3}^{(\phi^\dagger \phi,2)}$ to distinguish it from the vertex $\gamma^{(1,2)}$ discussed before \eqref{gluing4}, which in this subsection corresponds to $\gamma_{\mu\nu\alpha_2\alpha_3}^{(F_{\mu\nu} \phi^\dagger \phi,2)}$. The latter vertex identically vanishes, as mentioned before.}
Since $\gamma^{(1,3)}$ is manifestly $O(p_1)$ and we have already shown that  $\gamma_{\alpha_2\alpha_3}^{(\phi^\dagger \phi,2)}=O(p_2 p_3)$, we conclude that
\be
\gamma_{\mu\nu\alpha_1\alpha_2\alpha_3}^{(1,3)}(q_1=0,p_1,p_2,p_3) = O(p_1p_2p_3)\,,
\ee
as expected. We can see how in this case one power of $p_i$ comes directly from the Feynman rules of the composite operator vertex while the others emerge after the integration over the virtual momentum. 

\section{Gauge-variant operators and IR divergences}
\label{sec:gauge_variant}
We have proved in section \ref{sec:IRFG} that correlation functions of gauge invariant operators are IR finite. But the converse is not true, namely there exist correlation functions
of gauge-variant operators that are also IR finite. Indeed, we have already proved the IR finiteness of the $n$-point photon correlators $\langle A_{\mu_1}(p_1) \cdots A_{\mu_n}(p_n)\rangle $. In this case we can understand the reason of this property: the quantum corrections of the photon $n$-point functions can be related to the $n$-point functions of conserved currents $J_\mu$ which are indeed gauge invariant and then IR finite.
 
 Another set of gauge-variant correlators that we have implicitly proved to be IR finite during our discussion are $\langle \mathcal{O}_1(q_1) \cdots \mathcal{O}_k(q_k) A_{\mu_1}(p_1)\cdots A_{\mu_m}(p_m) \rangle$ where $\mathcal{O}_i$ are gauge-invariant operators. In this case the finiteness is a consequence of the relation (\ref{eq:gammaJi0}) since, as in the case of the connected correlator
 $\langle \mathcal{O}_1 \cdots \mathcal{O}_k\rangle$, also these correlators can be constructed using the $\gamma^{(\mathcal{O}_1 \ldots \mathcal{O}_k,k)}$ defined in section \ref{sec:IRFG} as building blocks. However these are not the only gauge-variant correlation functions that happen to be IR finite. In this section we will prove that another notable non-gauge invariant correlator is IR finite to all orders in perturbation theory: the two-point function of elementary matter fields
 $\langle \Phi^{\dagger}(q)\Phi(-q)\rangle $. In this case the IR finiteness is not a direct consequence of the equation (\ref{eq:gammaJi0}) and so the proof deserves a dedicated analysis.
 
 \subsection{IR finiteness of $\langle \Phi^{\dagger}(q)\Phi(-q) \rangle$} 
 In what follows we use the notation of section \ref{sec:IRFG} where with $\Phi$ we denote any elementary matter field that can be a boson or a fermion.
 
 In order to study this two-point function we can use the same idea introduced in section \ref{sec:IRFG} in which some building blocks are used in order to reconstruct the entire correlator. We define an effective action as in \eqref{eq:Seff} with $\mathcal{O}_1=\Phi$, $\mathcal{O}_2=\Phi^{\dagger}$, in terms of which the connected correlation function with $2k$ external matter fields and $n$ external non-dynamical photons is
\be
\widetilde{\gamma}_{\mu_1 \ldots \mu_n}^{(2k,n)}(x_i,y_j)\equiv  \Big(\!\prod_{i=1}^{k} \frac{\delta}{\delta J_{\Phi}(x_i)} \Big) \Big(\!\prod_{i=k+1}^{2k}\!\! \frac{\delta}{\delta J_{\Phi^{\dagger}}(x_i)} \Big) \Big(\!\prod_{j=1}^n \frac{\delta}{\delta A_{\mu_j}(y_j   )} \Big) S_{eff} \Big|_{A=J_{\Phi}=J_{\Phi^{\dagger}}=0}.
\ee 
As in the case of gauge-invariant correlation functions, the $n$-point functions of matter fields are constructed by gluing together the external photon lines using the vertices $\widetilde{\gamma}^{(2k,n)}$, although crucially we now have 
\be
p^{\mu_j}_j \widetilde{\gamma}_{\mu_1 \ldots \mu_n}^{(2k,n)} \neq 0\,, \quad  {\rm  for} \;\; k>0.
\ee
For small momenta $p_j$, we then have
\be
\widetilde{\gamma}_{\mu_1 \ldots \mu_n}^{(2k,n)}(q_i ,p_j) \not= O(p_1 \ldots p_n )\,, \;\; \quad  {\rm  for} \;\; k>0\,,
\label{eq:WIgaugevariant}
\ee
 while 
 \be
 \widetilde{\gamma}_{\mu_1 \ldots \mu_n}^{(0,n)}(p_j) \equiv \gamma_{\mu_1 \ldots \mu_n}^{(n)}(p_j)= O(p_1 \ldots p_n) \;.
 \ee
In what follows we will show that the missed powers of $p_i$ in (\ref{eq:WIgaugevariant}) are not sufficient to produce an IR divergence in the matter two-point function $\langle \Phi^{\dagger}(p)\Phi(-p) \rangle$. This is not the case for higher gauge-variant correlators, as already shown in figure \ref{fig:4fermion} for the matter $4$-point function.
 
 The correlator $\langle \Phi^{\dagger}(q)\Phi(-q) \rangle$ is necessarily composed by one, and only one, $\widetilde{\gamma}^{(2,n)}$ and an arbitrary number of $\gamma^{(m)}$, where $m \geq 2$ ($m=1$ would correspond to an insertion of a photonic tadpole which is zero by Lorentz invariance)  as shown in figure \ref{fig:2pt_gamma}. 
 It is useful to divide the set of all diagrams in two categories:
 \begin{figure}
\centering
\begin{subfigure}{.5\textwidth}
  \centering
  \includegraphics[width=.6\linewidth]{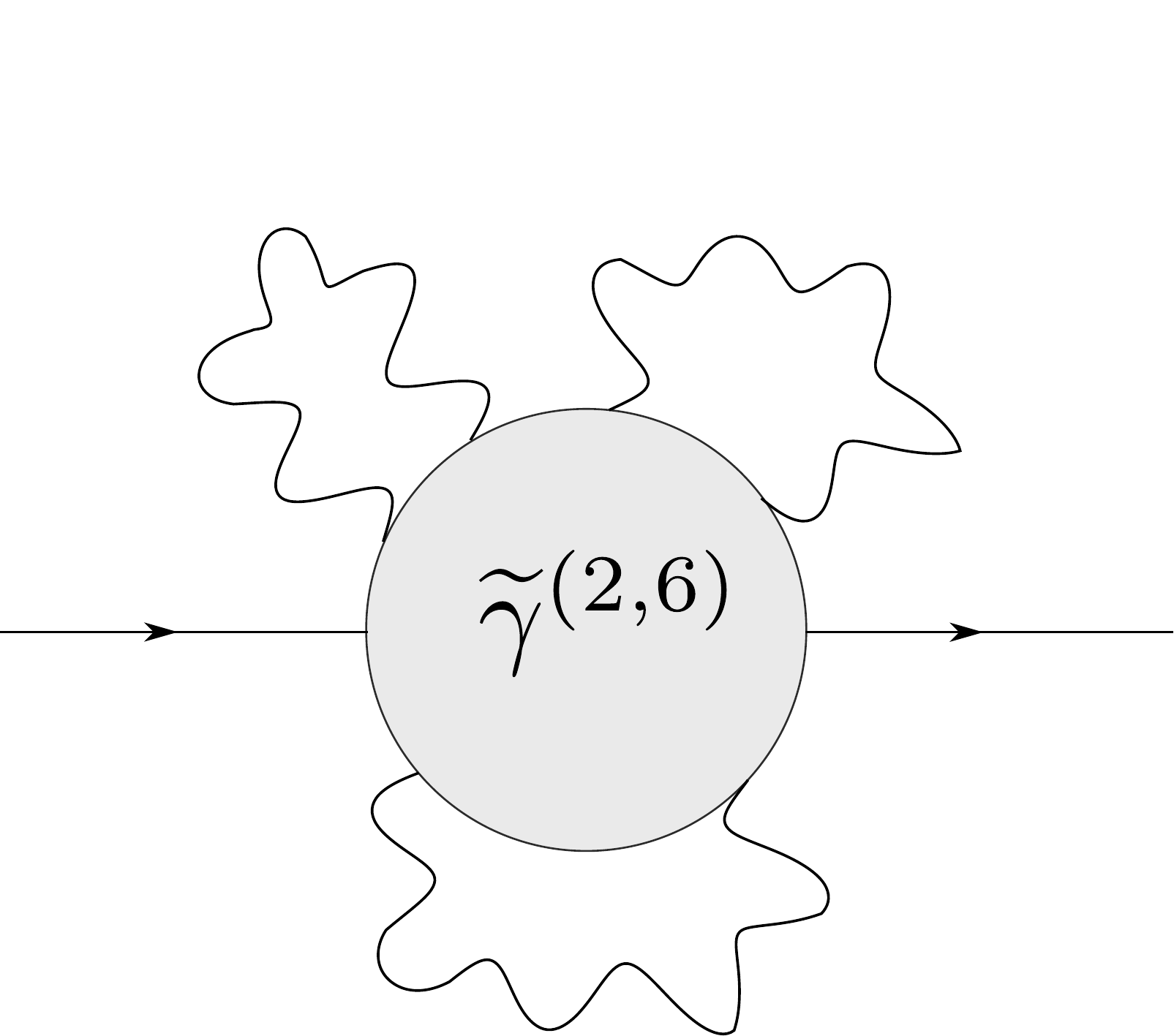}
  \caption{}
  \label{fig:typeA}
\end{subfigure}%
\begin{subfigure}{.5\textwidth}
  \centering
  \includegraphics[width=.6\linewidth]{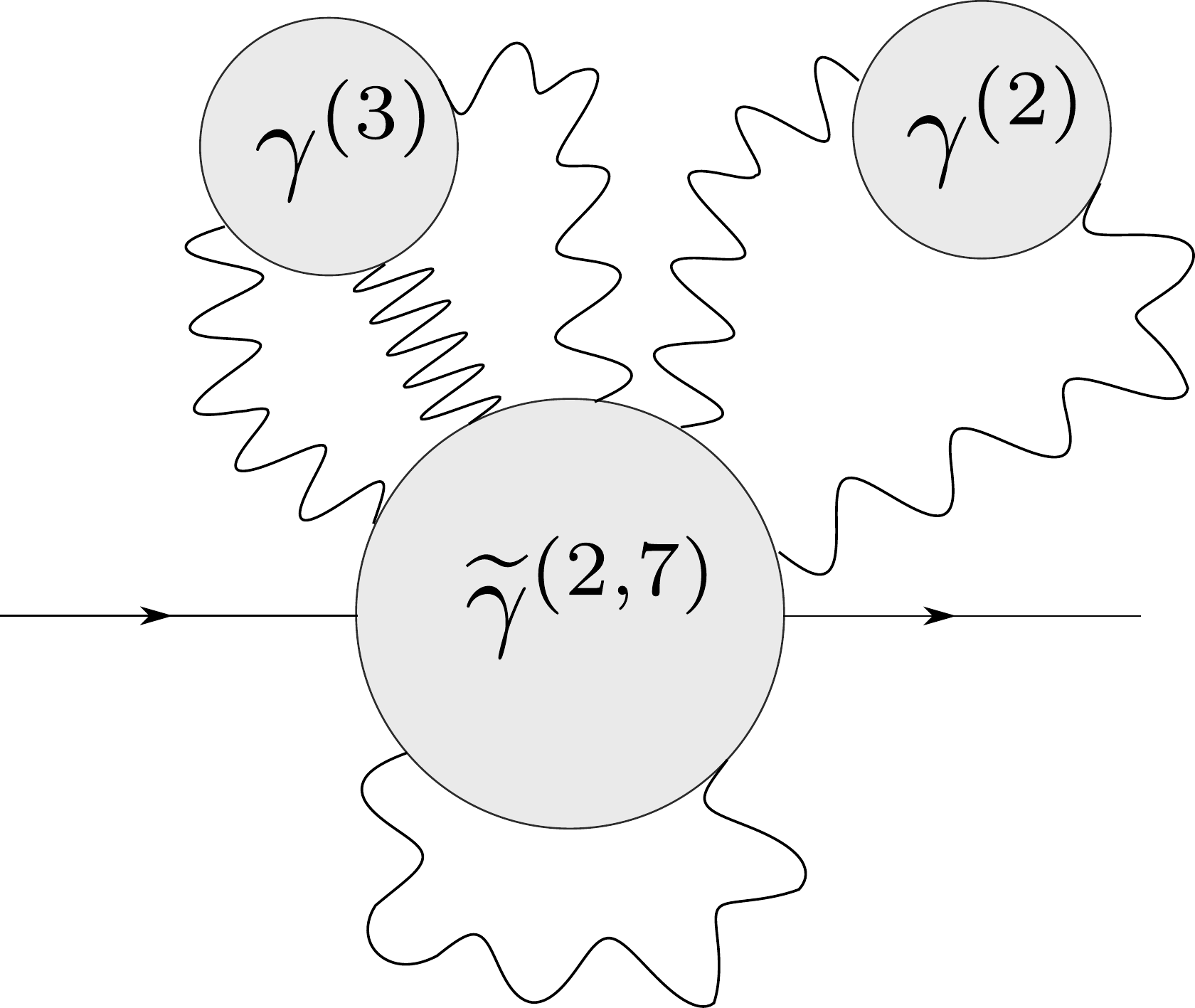}
  \caption{}
  \label{fig:typeB}
\end{subfigure}
\caption{Example of a type A (left) and a type B (right) diagram contributing to the two-point function $\langle \Phi^{\dagger}(q)\Phi(-q) \rangle$}
\label{fig:2pt_gamma}
\end{figure}

\begin{itemize}
    \item We call of type A all the graphs obtained by the ones entering  $\widetilde{\gamma}^{(2,2n)}$, where
     the $2n$ photon lines are connected between each other, for any $n$.
   \item We call of type B the graphs which are not of type A, namely the ones in which at least one $\gamma^{(m)}$ is present.
 \end{itemize}

Each type A diagram is individually IR finite. Indeed, for every graph ${\mathcal G}$ (both type A and type B) contributing to $\langle \Phi^{\dagger}(q)\Phi(-q) \rangle$, 
simple graph topology considerations give
\be
n \leq L  \,,
\label{relationLIgamma}
\ee
where $L$ represents the number of loops of $\mathcal{G}$ and $n$ is the number of internal photon lines.\footnote{In particular $L=n$ if we do not include matter self-interactions in the theory while $L \geq n$ if they are present.} By definition a type A graph $\mathcal{G}_A$ is $n$-photon particle irreducible.
 We can hence assign independent integrated momenta to each internal photon line, and thanks to \eqref{relationLIgamma} we always have sufficient independent momenta to do it. Then from a simple power counting we get that for small integrated momenta 
\be
\mathcal{G}_A \sim \int \frac{d^3p_1 \cdots d^3p_{L}}{p_1^2\cdots p_{n}^2} F(p_1, \cdots, p_{L})
\label{typeAint}
\ee
where $F$ is an analytic function in the origin of the momentum space since the matter is massive. The integral (\ref{typeAint}) is then IR convergent due to \eqref{relationLIgamma} 
for any choice of $\mathcal{G}_A$, at any loop order $L$.

Consider now the type B diagrams.  In this case the external photon lines of $\widetilde{\gamma}^{(2,n)}$ can be attached to another photon line of the same vertex or to another vertex $\gamma^{(m)}$, see the right panel in figure \ref{fig:2pt_gamma}. In the former case, doing an analysis similar to the one performed above for the type A diagrams, we conclude that no IR divergences can arise. The same is not true for the latter case. Individual IR divergences can now appear but their total contribution will be regulated by the powers of $p_i$ coming from $\gamma^{(m)}$,  thanks to gauge invariance that still ensures the validity of \eqref{eq:gammapzero}. For every block $\gamma^{(m)}$ we can assign $m-1$ independent momenta  $p_{1}\cdots p_{m-1}$ while the last one will be fixed by the conservation of the total momentum. Then from a simple power counting consideration we get that at small $p_i$ the contribution of the $m$ photons connecting $\widetilde{\gamma}^{(2,n)}$ to $\gamma^{(m)}$ is
\begin{equation}
    \int d^3p_1\dots d^3p_{m-1} \frac{p_1\dots p_m}{p_1^2\dots p_m^2}
\end{equation}
which is convergent for $m \geq 2$. Then we conclude that also the type B diagrams are IR finite at all orders in perturbation theory.
The argument above generalizes to correlators constructed using only one vertex $\widetilde{\gamma}^{(2k,n)}$ and an arbitrary number of $\gamma^{(n)}$. This is the case for instance of the $(n+2)$-point function  $\langle \Phi^{\dagger}(q_1)\Phi(q_2)A_{\mu_1}(p_1)\ldots A_{\mu_n}(p_n)\rangle$, which will also be IR finite to all orders in perturbation theory.

Matter $n$-point functions with $n\geq4$ are instead affected by uncancelled IR divergences. A generic diagram can now be composed of more than one vertex $\widetilde{\gamma}^{(2k,n)}$. By connecting two (or more) photons coming from these $\widetilde{\gamma}$'s, the poles $1/p_i^2$ are no longer suppressed by momentum factors 
and IR divergences can arise. For instance, we can interpret the graph in figure \ref{fig:4fermion} as the diagram composed of two (tree-level) vertices $\widetilde{\gamma}^{(2,2)}$ connected together. Then the pole $1/p^4$ generated from the photons at the exceptional incoming total momentum $Q = q_1+q_2 = 0$ is not regulated and an IR divergence is found.

 \section{The role of monopole operators}
 \label{sec:moop}

We have shown in this paper that correlation functions of gauge-invariant operators (and some gauge-variant ones) in 3d euclidean abelian gauge theories are IR finite when
matter fields are massive. Our results are general and apply independently of the nature of the theory: effective or fundamental. They are also insensitive to the global structure of the gauge group, i.e. $U(1)$ vs $\mathbb{R}$. At the non-perturbative level, however, things might change. In particular, the so called monopole operators can significantly affect the behaviour of the theory at low energies. In this section we would like to briefly review what is known about the role of monopole operators in order to put our results in a wider perspective and understand in which situations we might expect non-perturbative corrections to the euclidean correlation functions to be absent.

Abelian gauge theories in 3d can admit a trivially conserved topological $U(1)_{{\rm T}}$ global symmetry given by
\be
J_\mu^{{\rm T}} = \frac{1}{4\pi} \epsilon_{\mu\nu\rho} F^{\nu \rho}\,.
\label{eq:TopCurr}
\ee
The states charged under $U(1)_{{\rm T}}$ have magnetic charge and are denoted monopoles.
The local operators charged under $U(1)_{{\rm T}}$ are denoted monopole operators \cite{Borokhov:2002ib}.
In Euclidean space, monopoles are finite action configurations and should be regarded as instantons (for instance, the reduction to 3d of ordinary 4d monopoles), while in Minkowski space-time monopole operators create vortex configurations (for instance, the uplift to 3d of 2d instanton vortex configurations).
Using conventional terminology, we will denote them as monopoles.
Such states are quantum mechanically well defined only if the magnetic flux is quantized, namely if the gauge symmetry is globally a compact $U(1)$ and not $\mathbb{R}$. Monopoles can hence be present only for $U(1)$ gauge theories. 

The actual importance of monopoles for compact $U(1)$ gauge theories depends on how the $U(1)_{{\rm T}}$ symmetry is 
realized in the theory. From an RG point of view, it is useful to rephrase the impact of monopoles in terms of the scaling dimensions of the monopole operators  seen as deformations in the UV theory, see e.g. \cite{Hermele:2004hkd}.
If in the UV the $U(1)_{{\rm T}}$ is preserved, like in UV-complete 3d gauge theories in the continuum, monopole deformations can be forbidden by simply demanding $U(1)_{{\rm T}}$ conservation. In this case, however, monopole operators can still take a VEV and induce a spontaneous symmetry of $U(1)_{{\rm T}}$. Large $N$ considerations show that abelian gauge theories flow to an interacting CFT where $U(1)_{{\rm T}}$ is unbroken. For sufficiently low numbers of flavour a spontaneous ``chiral" symmetry or a first-order transition might occur in the fermion or boson cases respectively, where monopoles can condense. The spontaneous symmetry of $U(1)_{{\rm T}}$ would lead to a Goldstone-boson, which is the dual photon. In this case the theory will flow in the IR to a Coulomb free phase.

The situation is quite different when $U(1)_{{\rm T}}$ is explicitly broken in the UV and abelian gauge theories are only approximate effective descriptions at some energy scale. Spin systems on lattices and Polyakov's $SU(2)$ model \cite{Polyakov:1976fu} are notable examples in this class. Monopole deformations are now allowed. Their impact on the IR physics depend on whether they correspond to irrelevant or relevant deformations. At large $N$, monopole deformations in QED$_{3}$ scale as $N$ and are irrelevant \cite{Borokhov:2002ib}. The same applies to scalar ${\mathbb{CP}}^{N-1}$ models \cite{Murthy:1989ps} (see also \cite{Metlitski:2008dw} for a more modern analysis in terms of monopole operators), which are expected to describe the phase transition between the Neel and the valence bond state solid phases of certain anti ferromagnetic spin lattice systems \cite{PhysRevB.70.144407}. For sufficiently low number of flavours, monopoles can be relevant.\footnote{They could also be relevant in the IR, but irrelevant in the UV, i.e. in high energy parlance they could be dangerously irrelevant operators.} In the latter case they have been shown to lead to a confining gapped phase in the IR \cite{Polyakov:1975rs,Polyakov:1976fu}.\footnote{The appearance of a trivially gapped phase can often be ruled out by  't Hooft anomaly matching arguments \cite{Komargodski:2017dmc}.}

\section{Outlook}
\label{sec:outlook}
The main motivation of this paper was to investigate whether UV-complete, parity-invariant, abelian gauge theories at finite $N$ and at fixed dimension $d=3$
could in principle be studied in perturbation theory, like it has successfully been done with massive quartic vector models in both $d=3$ and $d=2$ for decades (see e.g. \cite{Pelissetto:2000ek} for a relatively recent review). Our work provides a first positive answer, as long as gauge-invariant correlators or specific gauge-variant correlators like the two-point function of elementary operators are used.\footnote{In sQED$_3$ one could get rid of IR divergences by studying the theory in the classically Higgsed phase (see e.g. \cite{Kleinert:2001hr}), though perturbation theory gets more complicated.} As we mentioned, if Chern-Simons terms are included, IR finiteness is trivially guaranteed.
But of course this is not the end of the story. 
As discussed in section \ref{sec:moop}, non-perturbative effects related to monopole operators are not expected to occur in UV-complete theories with unbroken $U(1)_{{\rm T}}$.  
But being the theories strongly coupled in the IR, a perturbative analysis would still be meaningless, unless a Borel resummation procedure is implemented.
In  contrast to the quartic vector models in both $d=3$ and $d=2$, there are no proofs about the Borel summability of the perturbative series in abelian 3d gauge theories.
We also do not have sharp predictions for the large order behaviour of the series, see \cite{LeGuillou:1990nq} for a review with references to early attempts in this respect.
Despite that, the prospects seem promising. 

Instanton and renormalon singularities are so far the only known obstructions to the Borel summability of perturbative series in QFT. Both QED$_3$ and sQED$_3$ (with no sextic interaction) are super-renormalizable theories with no marginal couplings,  for which no renormalon singularities are expected to appear. In QED$_3$ no instanton configurations can arise and we are not aware of known instanton configurations in sQED$_3$ on $\mathbb{R}^3$.
Let us then assume that perturbation theory can capture the long distance properties of these theories, provided a sufficient number of perturbative coefficients are known, so that
a sufficiently accurate Borel function can be numerically reconstructed.\footnote{It is also possible that the perturbative series are non-Borel resummable, yet perturbation theory
is able to capture non-perturbative physics by a resurgent analysis. In order to do so, however, a large number of perturbative coefficients would be needed.}
We could then consider a physical renormalization scheme and compute zeros of Borel resummed $\beta$-functions  (for sQED$_3$ this would require to define 
the quartic scalar self-interaction by means of gauge invariant correlators to avoid IR divergences) like it has been proposed long ago for quartic models in \cite{Parisi:1980gya}.
Alternatively, we could for example compute two-point functions of gauge invariant operators and see the evolution of the mass gap $M/m$ as a function of $e^2/m$ (for sQED$_3$ at fixed Ginzburg parameter $k$), where $m$ is a renormalized UV mass, as more recently done for both 2d  \cite{Serone:2018gjo} and 3d scalar quartic models \cite{Sberveglieri:2020eko}.
The last possibility seems more feasible, because it only requires the computation of two-point functions.

We believe an analysis of this kind (with or without Chern-Simons terms) would be useful to assess the existence of a critical behaviour in both sQED$_3$ and QED$_3$ at finite $N$, help us in finding the critical values $N_c$ where these theories exit their conformal windows, and possibly to provide concrete checks of $3d$ dualities between abelian gauge theories.

It would be interesting to investigate if gauge-invariant correlators in non-abelian euclidean $3d$ gauge theories are also IR finite. In this case a simple use of Ward-Takahashi identities is not available and one should probably undertake a more involved analysis using BRST symmetry.

\section*{Acknowledgments}

We thank Francesco Benini and Lorenzo Di Pietro for useful discussions. Work partially supported by INFN Iniziativa Specifica ST\&FI. 

\bibliographystyle{JHEP}
\bibliography{Refs}

\end{document}